\documentclass[aps,prd,twocolumn,superscriptaddress,groupedaddress, showkeys,floatfix]{revtex4} 
\usepackage{placeins}

\usepackage{subfig}
\usepackage{natbib}
\usepackage{mathtools}
\usepackage{scalerel}
\usepackage{graphicx}  
\usepackage{bm}        
\usepackage{tikz}
\usepackage{amssymb}   
\usepackage{lipsum}
\usepackage{caption}
\usepackage{lineno}
\usetikzlibrary{svg.path}
\usepackage{amsmath,array}
\usepackage{hyperref}
\hypersetup{
	colorlinks=true,
	linkcolor=blue,
	filecolor=magneta,      
	urlcolor=blue,
}
\DeclareCaptionJustification{justified}{\leftskip=0pt \rightskip=0pt \parfillskip=0pt plus 1fil}

\definecolor{orcidlogocol}{HTML}{A6CE39}
\tikzset{
    orcidlogo/.pic={
        \fill[orcidlogocol] svg{M256,128c0,70.7-57.3,128-128,128C57.3,256,0,198.7,0,128C0,57.3,57.3,0,128,0C198.7,0,256,57.3,256,128z};
        \fill[white] svg{M86.3,186.2H70.9V79.1h15.4v48.4V186.2z}
        svg{M108.9,79.1h41.6c39.6,0,57,28.3,57,53.6c0,27.5-21.5,53.6-56.8,53.6h-41.8V79.1z M124.3,172.4h24.5c34.9,0,42.9-26.5,42.9-39.7c0-21.5-13.7-39.7-43.7-39.7h-23.7V172.4z}
        svg{M88.7,56.8c0,5.5-4.5,10.1-10.1,10.1c-5.6,0-10.1-4.6-10.1-10.1c0-5.6,4.5-10.1,10.1-10.1C84.2,46.7,88.7,51.3,88.7,56.8z};
    }
}
\newcommand\orcidicon[1]{\href{https://orcid.org/#1}{\mbox{\scalerel*{
                \begin{tikzpicture}[yscale=-1,transform shape]
                \pic{orcidlogo};
                \end{tikzpicture}
            }{|}}}}

\begin{document}
\title{Dark ages bounds on non-accreting massive compact halo objects}

\author{Vivekanand Mohapatra$^{\orcidicon{0000-0002-5816-5225}}$\,}
    \email[Email address: ]{vivekanandmohapatra@gmail.com} 
    \affiliation{Department of Physics, National Institute of Technology Meghalaya, Sohra, Meghalaya, India}

\author{Alekha C. Nayak$^{\orcidicon{0000-0001-6087-2490}}$\,}
    \email[Email address: ]{alekhanayak@nitm.ac.in}
    \affiliation{Department of Physics, National Institute of Technology Meghalaya, Sohra, Meghalaya, India}

\begin{abstract}
We derive a complementary cosmological upper bound on the fraction of dark matter residing inside non-accreting massive compact halo objects (MACHOs) using the cosmic dawn and dark ages global 21-cm signal $(T_{21})$. MACHOs of mass $M\gtrsim 10^3~M_\odot$ moving through the baryonic fluid during post-recombination transfer kinetic energy to the intergalactic medium via dynamical friction, thereby raising the gas temperature and distorting the 21-cm signal predicted in the $\Lambda$CDM framework. We consider both a monochromatic distribution of MACHOs and two extended mass distributions: log-normal and critical collapse. Imposing the conditions that the deviation in the global 21-cm signal $\Delta T_{21}$ does not exceed $50~\rm mK$ at $z\sim 17$ or $15~\rm mK$ at $z\sim 89$, and that no emission signal appears at $z \gtrsim 300$, we derive upper bounds on the MACHO fraction $f_M$ across the mass range $10^3 \lesssim M_c/M_\odot \lesssim 10^7$. The dark ages and $z \gtrsim 300$ criterion yield constraints that are both tighter and free from astrophysical uncertainties associated with star formation, particularly $f_M = 0.03$ for $M_c = 10^7~M_{\odot}$, providing a complementary cosmological window. The extended mass distributions yield bounds that are less stringent than those of their monochromatic counterparts.

\end{abstract}

\maketitle

\section{Introduction}
\label{sec:intro}
The particle nature of dark matter (DM) remains a cornerstone in the field of particle physics and cosmology. DM is often considered collisionless and cold in a $\Lambda$CDM framework \cite{Geller:1989da,2DFGRS:2001zay, SDSS:2000hjo, Springel:2005nw}. However, alternative cases such as fuzzy and warm DM \cite{Matos:1998vk, Hu:2000ke, Arbey:2001qi, Hui:2016ltb, Boyle:2001du, Chatterjee:2019jts}, and self-interacting DM \cite{Carlson:1992fn,Spergel:1999mh,Bhatt:2019qbq} have also been proposed. Apart from particle DM with masses in the GeV and sub-GeV range, dark matter may also consist of compact and massive objects \cite{Green:2020jor}. For instance, enhanced primordial power spectrum at small scales from inflation \cite{Leach:2001zf, Pajer:2013fsa, Hertzberg:2017dkh, Mishra:2019pzq, Inomata:2021tpx, Hooper:2023nnl}, reheating \cite{Erickcek:2011us, Erickcek:2015jza}, supersymmetry \cite{Fan:2014zua}, or a codecaying dark matter scenario \cite{Dror:2017gjq},
can produce objects of varying density and mass distributions. These massive objects present in beyond-the-Standard Model scenarios include quark nuggets \cite{Witten:1984rs}, axion stars \cite{Ruffini:1969qy, Kolb:1993zz, Liebling:2012fv, Chang:2024fol}, Q-balls \cite{Coleman:1985ki}, non-topological solitons \cite{Lee:1991ax}, dark blobs \cite{Grabowska:2018lnd}, and primordial black holes \cite{Green:2020jor, Carr:2020xqk}--- often referred to as massive compact halo objects (MACHOs) in literature.

In most scenarios, MACHOs are assumed to interact gravitationally with baryons. Moreover, MACHOs with masses $M>10^{-10}~M_\odot$ that make up all the dark matter are heavily constrained \cite{Graham:2023unf}. A more generic constraint is obtained from gravitational microlensing by analysing MACHO collaboration \cite{Macho:2000nvd}, EROS-2 \cite{EROS-2:2006ryy}, and OGLE-IV \cite{Udalski2015OGLEIVFP, Mroz:2024mse, Mroz:2024wia} data for mass $M\lesssim \mathcal {O}(10)~M_\odot$ \cite{Blaineau:2022nhy}. For mass ranges $M\gtrsim \mathcal {O}(10)~M_\odot$, the primary constraints are obtained from gravitational lensing \cite{Macho:2000nvd, Niikura:2019kqi, EROS-2:2006ryy, Oguri:2017ock, Wilkinson:2001vv}, dynamical heating of interstellar medium and stars \cite{Lu:2020bmd, Takhistov:2021aqx, Brandt:2016aco, Wadekar:2022ymq, Graham:2023unf, Graham:2025opw}, and accretion \cite{Bai:2020jfm}. There are also primordial black holes with specific constraints emerging from the detection of gravitational waves \cite{Macho:2000nvd, Niikura:2019kqi, EROS-2:2006ryy, Oguri:2017ock, Wilkinson:2001vv}, modelling the accretion of gas \cite{Moore:1993sv, Yoo:2003fr, Carr:2020gox, Ramirez:2022mys, Kohri:2022wzp}, and from the impact on the cosmic microwave background \cite{Ricotti:2007au, Ali-Haimoud:2016mbv}.
Although microlensing and dynamical heating provide compelling probes of MACHOs, they also carry their own uncertainties. This makes a complementary cosmological probe valuable to limit the abundance of MACHOs as a subcomponent of DM.

The 21-cm hyperfine transition of neutral hydrogen provides a powerful cosmological probe of the post-recombination intergalactic medium. In the $\Lambda$CDM framework, the differential brightness temperature is expected to show two distinct absorption troughs: one during the dark ages $(z \sim 89)$, and the other during the cosmic dawn $(z \sim 17)$ \cite{Pritchard:2005an, Pritchard:2011xb, Furlanetto:2006jb, Furlanetto:2006tf}. The EDGES experiment has reported a tentative absorption feature centred at $z \sim 17.2$ with an amplitude of $-0.5^{+0.3}_{-0.5}~\rm K$ \cite{Bowman:2018yin}, although the SARAS 3 experiment has rejected the EDGES signal at 95\% CL \cite{Singh:2021mxo}, and the tension remains unresolved. Future experiments such as REACH \cite{deLeraAcedo:2022kiu}, FARSIDE \cite{2019BAAS...51g.178B}, SEAMS \cite{10.1063/5.0043435}, PRATUSH \cite{2023ExA....56..741S}, and LuSee-Night \cite{2023arXiv230110345B}, are expected to shed light on this discrepancy. Unlike the cosmic dawn signal, which is sensitive to uncertain astrophysical processes such as star formation efficiency, X-ray heating, and Lyman-$\alpha$ emissivity, the dark ages signal at $z \gtrsim 30$ is free from these uncertainties — the universe remains largely homogeneous and isotropic in the absence of any astrophysical structure during this epoch \cite{Pritchard:2005an, Pritchard:2011xb, Furlanetto:2006jb, Furlanetto:2006tf}. Therefore, any deviation from the standard dark ages signal would be a clear indicator of nonstandard physics. The recent
proposal for LuSee-Night to reach the farside of the Moon in 2026 aims to observe the sky in the frequency range of $0.1-50$ MHz, which may allow for the detection of the global 21-cm signal from the dark ages \cite{10906958}. Moreover, for future lunar-based experiments, an integration time of 20,000 h is anticipated to achieve an uncertainty $(\Delta T_{21})$ of $15$ mK in detecting the standard dark ages signal. Furthermore, extending the integration time to 100,000 h can reduce the uncertainty to $5$ mK \cite{Burns:2020gfh, Rapetti:2019lmf}.

In this work, we consider MACHOs of masses $M\gtrsim10^3~M_\odot$ interacting gravitationally with baryonic fluid in the post-recombination era. These objects can induce wakes, and the relative motion between the object and the wake produces a frictional force. This frictional force can dissipate energy into the intergalactic medium via momentum transfer, heating the gas. In the previous work \cite{Bhalla:2025orw}, the authors studied the thermal evolution of the intergalactic medium and its impact on the cosmic dawn signal. We extend this study further by evaluating the thermal evolution of gas in the presence of this energy dissipation and its impact on the dark ages global 21-cm signal, along with the cosmic dawn. Additionally, we consider extended mass distributions along with the monochromatic distribution of these objects. We find that the dark ages can provide stronger and astrophysical uncertainty-free constraints, while probing a wider mass window.

The rest of the paper is organised as follows: In Sec. \ref{sec:dynamical heating} we describe the formulation of dynamical heating induced by MACHOs, and in Sec. \ref{sec:extended_mass} we describe the extended mass spectrum considered in this work. Sec. \ref{sec:global_21_signal} describes the formulation of the cosmic dawn and dark ages global 21-cm signals. In Sec. \ref{sec:Therma_evolution} we study the thermal and ionisation evolution of the universe in the presence of dynamical heating. We present our findings by constraining the fraction of dark matter in Sec. \ref{sec:Result} and concluding remarks in Sec. \ref{sec:conclusion}.

\section{Dynamical heating and mass spectrum of MACHOs}
In this section, we discuss the effects of MACHOs on the IGM arising from their relative motion during the post-recombination era. Before proceeding, we clarify that the MACHOs considered in this work are not dense enough to accrete matter. For instance, extremely diffused axion stars and axion miniclusters with a radius greater than thrice their Schwarzschild radius are inefficient at accreting matter, and are constrained solely via dynamical heating in an astrophysical scenario \cite{Kim:2025gck}. Thus, our analysis considers that the energy dissipation into the IGM is solely due to dynamical heating--- as discussed later in this section. 

\subsection{Dynamical heating}\label{sec:dynamical heating}

During the pre-recombination era, baryons and radiation were tightly coupled via efficient Thomson scattering, and the baryonic sound speed $(c_s)$ evolves as $\sim c/\sqrt{3(1+R)}$, where $c$ and $R \equiv 3\rho_b/4\rho_\gamma$ represent the speed of light in vacuum and the baryon-to-photon ratio \cite{Tseliakhovich:2010bj, Scheck:2014cba}. As the universe expands and cools, $R(z)$ increases, leading to a decrease in $c_s$. Consequently, the baryons kinetically decouple from the radiation drag at redshift $z_d\sim 1060$, slightly after the recombination epoch \cite{Tseliakhovich:2010bj}. Soon after this, the formation of neutral hydrogen atoms reduces $c_s$ even further, making the IGM evolve as an ideal gas following the relation \cite{Tseliakhovich:2010bj}

\begin{equation}
    c_s = \sqrt{\frac{\gamma~k_BT_{gas}}{\mu m_H}}~,
\end{equation}
where $T_{gas}$ represents the kinetic temperature of the baryons, $\gamma = 5/3$ for an ideal monatomic gas, $\mu = 1.22$ is the mean molecular weight including a helium mass fraction of $0.24$, and $m_H$ is the mass of the hydrogen atom. Before recombination, the baryon-dark matter motion remains in a subsonic regime, quantified by the Mach number $\mathcal{M}\equiv v_{bc}/c_s$, where $v_{bc}$ is $10^{-5}$c and $c_s\sim c/\sqrt{3}$. However, soon after the recombination, the baryonic sound speed attains a value of $\sim 6~\rm km~s^{-1}$, while the dark matter velocity $(v_{bc})$ remains unaffected and scales as $(1+z)$ \cite{Hu:1994uz, Ma:1995ey, Tseliakhovich:2010bj}. During this era, the motion enters the hypersonic regime with $\mathcal{M}\sim 5$. The relative velocity $v_{bc}$ follows a Boltzmann distribution with an rms value of $\sim 30~\rm km~s^{-1}$, and remains spatially coherent on scales up to $\sim 3$ Mpc \cite{Silk:1967kq, Tseliakhovich:2010bj, Barkana:2010zq}. Throughout this work, we adopt the rms value as a representative bulk velocity for the dynamical friction calculation.

An object of mass $(M_{c})$ streaming through at velocity $(v_{bc})$ inside a uniform gaseous medium of average density $(\rho_b)$ can produce a density perturbation due to its gravitational potential, producing an overdense region known as wakes. In a linear regime, the density perturbation evolution follows an inhomogeneous wave equation, and the solution to it provides the response of the medium to a moving gravitational source, with the wake and Mach cone in the supersonic regime. For a more detailed explanation, we refer the reader to Ref. \cite{Ostriker:1998fa} for a linear analysis and Ref. \cite{Kim_2009} for a non-linear analysis of dynamical friction. The interaction between the point mass and wake is purely gravitational, and the kinetic energy of the perturber is lost to the gas medium via the drag force $F_{\rm DF}$. The drag force associated with the dynamical friction can be expressed as \cite{Ostriker:1998fa, 1986A&A...164..337J, 1990A&A...232..447J, Kim:2025gck, Bhalla:2025orw}

\begin{equation}
    \label{eq:dynamical_force}
    F_{\rm DF} = \frac{4\pi (GM)^2\rho_b}{v_{bc}^2}~\mathcal{A}(\mathcal{M, S})~,
\end{equation}
where $G$ represents the gravitational constant. $\mathcal{A}$ term regulates the integration limits and can be interpreted as a geometric factor which depends on the Mach number and Coulomb logarithm factor $\cal S$. Following the \cite{Kim:2025gck, Bhalla:2025orw}, we adopt an expression for $\cal A$ which is valid for all $\cal M$ defined piecewise as 

\begin{equation}
I_1 = \frac{1}{2}\ln\left[\frac{1+\mathcal{M}}{1-\mathcal{M}}\right] - \mathcal{M},
\quad \mathcal{M} < 1 - x_{\min},
\label{eq:I1}
\end{equation}
\begin{equation}
I_2 = \frac{x_{\min}}{4} - \frac{\mathcal{M}}{2}
      - \frac{1-\mathcal{M}^2}{4 x_{\min}}
      + \frac{1}{2}\ln\left[\frac{1+\mathcal{M}}{x_{\min}}\right],
\label{eq:I2}
\end{equation}
\begin{equation}
I_3 = -\frac{1}{4 x_{\min}}
      + \frac{(\mathcal{M} - x_{\min})^2}{4 x_{\min}}
      + \frac{1}{2}\ln\left[\frac{1+\mathcal{M}}{x_{\min}}\right],
\label{eq:I3}
\end{equation}
\begin{equation}
I_4 = \frac{1}{2}\ln\left[\frac{\mathcal{M}+1}{\mathcal{M}-1}\right]
      + \ln\left[\frac{\mathcal{M}-1}{x_{\min}}\right],
\quad \mathcal{M} > \mathcal{M}_{+},
\label{eq:I4}
\end{equation}
where Eq.~\eqref{eq:I2} holds for $1 - x_{\min}< 
\mathcal{M} < \mathcal{M}_{-}$, Eq.~\eqref{eq:I3} for
$\mathcal{M}_{-} < \mathcal{M} < \mathcal{M}_{+}$,
with $\mathcal{M}_{-} \equiv \sqrt{1 + x_{\min}^2}$,
$\mathcal{M}_{+} \equiv 1 + x_{\min}$, and
$x_{\min} \equiv (1 + \mathcal{M})/\cal S$. The factor $\cal A$ is given by $I_1$ in the deep-subsonic regime $(\mathcal{M} < 1 - x_{\min})$, by $I_2$ and $I_3$ in the near-sonic transition, and by $I_4$ in the supersonic regime $(\mathcal{M} > 1 + x_{\min})$. In general, the Coulomb logarithm factor $\ln(\mathcal{S}) \equiv \ln(r_{\rm max}/r_{\rm min})$ depends on the radius of perturber $(r_{\rm min})$ and the distance it travels over time $t$ \cite{Ostriker:1998fa}. 

We set $r_{\rm min} = GM/(c_s^2+v_{bc}^2)$ to the Bondi-Hoyle-Lyttleton (BHL) radius for MACHOs \cite{Hoyle_Lyttleton_1939, Bondi:1944rnk}. Within this radius, baryons are gravitationally captured by the MACHO and no longer contribute to the large-scale wake responsible for dynamical friction. Following Ref. \cite{Suzuguchi:2024btk}, we treat this as the minimum scale below which the linear drag formula of Eq. \eqref{eq:dynamical_force} is not applicable, effectively suppressing friction in the high-velocity regime. The drag force defined in Eq.~\eqref{eq:dynamical_force} is self-consistent when the induced wake originates from a linear perturbation \cite{Kim_2009}. The linear solution predicts a fractional overdensity at a distance $r$ from the perturber of
\begin{equation}
    \frac{\delta \rho_b}{\rho_b} \simeq \frac{G M}{c_s^{2}\, r}
    = \frac{r_B}{r},
    \label{eq:overdensity}
\end{equation}
where $r_B \equiv GM/c_s^{2}$ is the Bondi radius \cite{Ostriker:1998fa,Kim_2009}, so the wake of a point-like perturber is intrinsically nonlinear at $r \lesssim r_B$, irrespective of its mass. The authors in Ref. \cite{Kim_2009} quantified the nonlinear response using hydrodynamic simulations for a non-accreting Plummer sphere of softening radius $r_s$ and found that it is governed by a single parameter
$\eta \equiv r_B/[r_s(\mathcal{M}^{2}-1)]$. Note that we consider the same boundary condition for MACHOs here, which interact with baryons purely gravitationally. The peak wake overdensity $\sim r_B/r_s$ is weighted by the kinetic energy in excess of sonic. For $\eta \gtrsim 2$, a bow shock detaches from the perturber and stands at a distance

\begin{equation}
    r_{\rm sh} = \frac{GM}{c_s^{2}(\mathcal{M}^{2}-1)}
               = \frac{GM}{v_{bc}^{2}-c_s^{2}},
    \label{eq:standoff}
\end{equation}
behind which the gas settles into a quasi-hydrostatic envelope. The envelope is nearly spherically symmetric and pulls the perturber equally forward and backwards, thereby exerting no net drag, so the friction arises solely from the quasi-linear wake beyond $r_{\rm sh}$ (see Figs.~9 and 16 of Ref.~\cite{Kim_2009}). As the $r_{\rm sh}$ remains independent of $r_s$, the point-mass limit is well defined, and the
nonlinear drag reduces to the linear formula with the inner cutoff raised to the envelope scale. We therefore replace $r_{\rm min}$ in the supersonic
regime ($\mathcal{M}>1$) with

\begin{equation}
    r_{\rm min} = \frac{2\,GM}{v_{bc}^{2}-c_s^{2}}
    = r_{\rm BHL}\,\frac{2(\mathcal{M}^{2}+1)}{\mathcal{M}^{2}-1},
    \label{eq:rmin_nl}
\end{equation}

where $r_{\rm BHL} = GM/(c_s^2+v_{bc}^2)$ is the Bondi-Hoyle-Lyttleton (BHL) radius \cite{Hoyle_Lyttleton_1939, Bondi:1944rnk}, and the coefficient of two is calibrated to reproduce the suppression on the dynamical force $F_{\rm DF}(\eta/2)^{-0.45}$, as measured in Ref.~\cite{Kim_2009} across their simulated range $2<\eta<100$. This suppression shows that the drag in a subsonic region remains unaffected by nonlinearity \cite{Kim_2009}, and Eq.~\eqref{eq:I1} is retained.  Finally, when $r_{\rm min} \geq r_{\rm max}$, the envelope engulfs the entire
wake, and the perturbative description fails altogether; thus setting
$\mathcal{A}=0$, which governs the behaviour of the heaviest masses. Furthermore, we consider $r_{\rm max} = \mathcal{M}\lambda_J$, where $\lambda_J = \sqrt{\pi c_s^2/G\rho_b}\sim c_s/H$ is the Jeans length and $H$ represents the Hubble parameter \cite{Bhalla:2025orw}. Since $r_{\rm max}\sim v_{bc}/H$ represents the maximum comoving distance travelled by a MACHO over a Hubble time, and this remains well below the $\sim 3$ Mpc coherence scale of $v_{bc}$ at all redshifts of interest, the relative velocity field can be treated as uniform over the interaction volume.

Now, as the perturber traverses the gaseous medium, the relative velocity $v_{bc}$ decreases due to the drag force and its evolution can be expressed as

\begin{equation}
\label{eq:velocity_evolution}
    \frac{dv_{bc}}{dz} = \frac{v_{bc}}{(1+z)} + \frac{F_{\rm DF}}{(1+z)HM}~,
\end{equation}
where the first term on the right-hand side of the above equation represents a decrease in $v_{bc}$ due to Hubble expansion, and the second term represents deceleration over Hubble time due to the drag. Furthermore, the energy dissipated per unit time due to the dynamical friction is defined as $v_{bc}F_{\rm DF}$. We can write the volumetric energy dissipation rate for a spectrum of MACHOs as \cite{Kim:2005ma, Conroy:2007ps}

\begin{equation}
\label{eq:energy_dissipation}
    \frac{dE}{dVdt} = \int dM \frac{\rho_{dm}}{M} \Psi_M v_{bc}F_{\rm DF}
\end{equation}
where $\rho_{dm}$ is the dark matter density. For a monochromatic spectrum $\Psi_M  = f_{\rm M} \delta_D(M-M_c)$, with $f_M$ being the fraction of dark matter residing inside MACHOs, and $\delta_D$ is the Dirac-delta function. We define this fraction as $f_M \equiv \Omega_{\rm MACHO}/\Omega_{ dm}$, where $\Omega_{\rm MACHO}$ and $\Omega_{ dm}\approx 0.261$ represent the densities in the units of critical density. It is worth emphasising that, in addition to the monochromatic distribution, the mass spectrum may follow different extended mass distributions, as discussed below.

\subsection{Extended mass spectrum of MACHOs}\label{sec:extended_mass}

If the MACHOs span extended mass spectra, the number density of MACHOs in a range $(M, M+dM)$ is often expressed as $\Psi\equiv f_M~dn/dM$, such that $\int dM\Psi_M$ provides the fraction of dark matter residing inside MACHOs. Note that $\Psi_M$ is a distribution function and has units $[\rm mass]^{-1}$. In this work, we consider two types of distribution functions--- log-normal and critical collapse. Although these mass distributions are often studied in the literature to characterise the formation and distributions of primordial black holes (PBHs), we adopt this formalism in our study, considering that PBHs are a class of MACHOs.
\begin{figure}
    \centering
    \includegraphics[width=\linewidth]{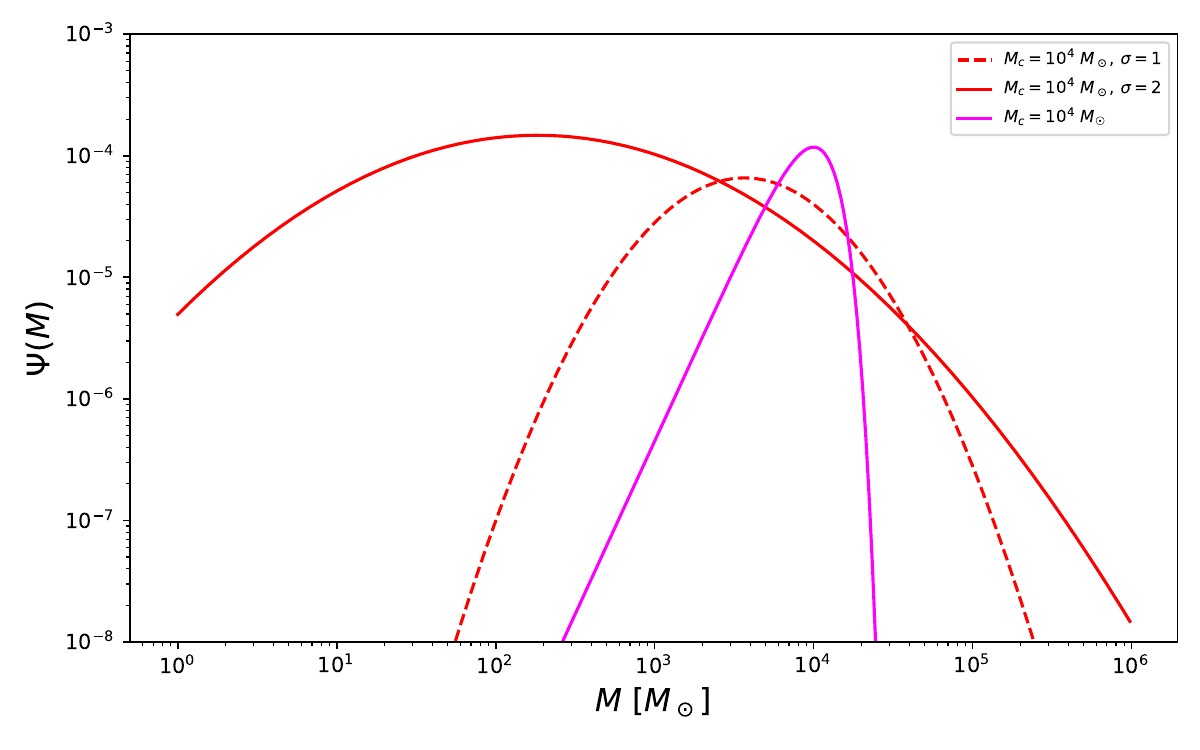}
    \caption{Shows the extended distribution function with respect to their masses. The dashed and solid red lines represent log-normal distributions for $M_c = 10^4M_{\odot}$ with $\sigma = 1$ and $2$, respectively. The magenta solid line represents the critical collapse distribution with the peak mass $10^4M_{\odot}$. 
    }
    \label{fig:extended_mass}
\end{figure}

\begin{enumerate}
    \item \textbf{Log-normal model:}
    In a log-normal model, the mass distribution function is expressed as \cite{Kannike:2017bxn, Green:2016xgy, Kuhnel:2017pwq, Dolgov:1992pu,  Carr:2017jsz}

    \begin{equation}
        \Psi_M = \frac{f_M}{\sqrt{2\pi}\sigma M}\exp\left\{ - \frac{\log^2[M/M_c]}{2\sigma^2}\right\},
    \end{equation}
    where $e^{-\sigma^2}M_c$ is the mass at which the distribution peaks, and $\sigma$ represents the width of the distribution. For illustration, we present the variation of $\Psi_M$ with respect to mass $M$ for $M_c = 10^4~M_{\odot}$ in the units of solar mass $(M_{\odot})$ in Fig. \ref{fig:extended_mass}. The dashed and solid red lines represent $\Psi_M$ for $\sigma = 1$ and $2$, respectively.

    \item \textbf{Critical collapse model:}
    The mass spectrum for the critical collapse model is expressed as \cite{Yokoyama:1998xd, Niemeyer:1999ak, Musco:2012au, Carr:2016hva, Carr:2017jsz}

    \begin{equation}
    \label{eq:critical_collapse}
        \Psi_M = f_M\frac{3.2}{M_c}\left(\frac{M}{M_c}\right)^{2.85}\exp\left[-\left(\frac{M}{M_c}\right)^{2.85}\right],
    \end{equation}
    where $M_c$ represents the mass at which $\Psi_M$ peaks. In Fig.~\ref{fig:extended_mass}, the magenta line presents the distribution for $M_c=10^4~M_{\odot}$. It can be observed that in the case of a log-normal distribution with $\sigma = 1$, the number density of MACHOs with mass $M_c$ is slightly lower than that following the critical collapse model. The impact of dynamical energy dissipation on the baryon temperature evolution for such a scenario will be discussed in Sec. \eqref{sec:Result}.

\end{enumerate}

Note that both the log-normal and critical collapse distribution functions are normalised for the mass range $0\to\infty$. This would translate to considering the existence of MACHOs with masses $0$ and $\infty$, which is apparently unphysical. On the other hand, if one considers the integration limits between, say, $(M_{\min}, M_{\rm max})$, the $f_M$ value cannot be fully recovered and remains sensitive to the integration limits. We can also observe that, for a given $M_c$ approximately $3\sigma$ masses remain within $[0.10, 10.24] M_c$ and $[0.32, 1.80] M_c$, for log-normal with $\sigma = 1$ and critical collapse distributions, respectively. On increasing $\sigma$ to 2, the $3\sigma$ range of masses for the log-normal distribution changes to $[0.01, 104.87] M_c$. Keeping this in mind, we considered $(M_{\min}, M_{\rm max})$ to be $(1, 10^9)~M_{\odot}$ as the integration limit to perform a conservative evaluation. The reason for considering these specific values can be interpreted by analysing the effect of dynamical friction on the thermal evolution of the IGM, which suggests that masses beyond $10^3\lesssim M_c/M_{\odot}\lesssim 10^7$ leave little to no signature on the IGM (see Sec. \ref{sec:Result} for a detailed explanation).

Now, the deceleration term $(a_{\rm DF} = F_{\rm DF}/M)$ shown in equations \eqref{eq:velocity_evolution} and \eqref{eq:energy_dissipation} is defined for a monochromatic distribution of MACHOs, which inherently fixes the masses of all MACHOs. However, MACHOs carry different number densities for different masses in an extended distribution scenario. Consequently, each MACHO of mass $M_i$ will decelerate at a rate $a_{\rm DF}(M_i)$ and dissipate the energy at redshift $z(M_i)$. Therefore, we need to track the velocity evolution of each MACHO to calculate the volumetric energy dissipation term.
In the further sections, we discuss the formalism of the global 21-cm signal and the impact of dynamical energy dissipation on the thermal evolution of baryons.

\section{21-cm signal}\label{sec:global_21_signal}
The post-recombination era $(z\lesssim1100)$ predominantly consists of neutral hydrogen atoms (HI), and residual electrons and protons. The hyperfine transition between the singlet $(F= 0)$ and triplet $(F= 1)$ states in HI can produce photons of wavelength $21$-cm $(\sim1420~\rm MHz)$. Observing the redshifted 21-cm signal can provide pristine cosmological and astrophysical information. Following the Rayleigh-Jeans limit, the 21-cm signal is often defined as the differential brightness temperature, expressed as
$T_{21} = [(T_s-T_{\gamma})/(1+z)]\, [1-e^{-\tau_{21}}]$. Here, $T_\gamma = T_{\gamma,0}(1+z)$
represents the CMB temperature with $T_{\gamma,0} = 2.725~\rm K$ at the present-day temperature. $T_s$ represents the spin temperature that determines the relative population
density of HI, and $\tau_{21}$ is the optical depth of 21-cm photons. In the limit $\tau_{21}\ll 1$, the differential brightness temperature or the global 21-cm signal can
be expressed as \cite{DAmico:2018sxd, Mitridate:2018iag, Cyr:2023iwu, PhysRevD.110.123506}

\begin{alignat}{2}\label{eq:T21}
     T_{21} \approx 27\mathrm{mK}~x_{\rm HI}\left(\frac{0.1424}{\Omega_mh^2}\right)^{0.5}
     \left(\frac{\Omega_bh^2}{0.0224}\right)& \left(1-\frac{T_\gamma}{T_s}\right) \nonumber \\ & 
     \sqrt{\frac{1+z}{10}}\, ,
\end{alignat}
where $x_{\rm HI}=n_{\rm HI}/{n_H}$ is the neutral hydrogen fraction, $n_H$ is the hydrogen number density, and $n_{\rm HI}$ is the neutral hydrogen number density. The evolution of $T_s$ depends on the interplay between the CMB, the baryonic temperature
$(T_{\rm gas})$, and Lyman alpha fields $(T_\alpha)$
\cite{Venumadhav:2018uwn, Furlanetto:2006tf, Acharya:2022txp}
\begin{equation}
    T_s^{-1} = \frac{T_\gamma^{-1}+x_cT_{\rm gas}^{-1}+x_\alpha T_\alpha^{-1}}{1+x_c+x_\alpha}\, ,
    \label{eq:Ts}
\end{equation}
where $x_c$ and $x_\alpha$ are the collisional and Wouthuysen-Field coupling coefficients,
respectively \cite{1952AJ.....57R..31W, 1959ApJ...129..536F, 1958PIRE...46..240F}. From
Eq.~\eqref{eq:T21}, we can observe that an absorption (emission) signal is expected when $T_s< T_\gamma~(T_s>T_\gamma)$. Below, we briefly discuss the two absorption troughs expected in the $\Lambda\rm CDM$ framework.

\subsection{The dark ages signal}

After recombination, the IGM and CMB are tightly coupled via inverse Compton scattering
between the CMB photons and the residual free electrons/protons. In the absence of any
Lyman alpha sources, $T_s$ follows the CMB temperature, making $T_{21}$ effectively zero
[see Eq.~\eqref{eq:Ts}]. However, the Compton scattering rate becomes ineffective
$\Gamma_c\ll H$, where $H$ represents the Hubble rate at redshifts $z\lesssim200$,
allowing the CMB and gas temperature to evolve as $(1+z)$ and $(1+z)^2$, respectively.
The collisional coupling remains effective at redshifts $40\lesssim z\lesssim 200$, where
$T_s$ remains coupled to $T_{\rm gas}$, leading to an expected absorption signal of
amplitude $\sim 42~\rm mK$ at $z\simeq 89$. The collisional coupling can be expressed
as \cite{2001A&A...371..698L, 2006ApJ...637L...1K, Furlanetto:2006tf, Pritchard:2011xb},
\begin{equation*}
    x_c = \frac{T_*}{T_\gamma}\,\frac{n_ik^{i\rm H}_{10}}{A_{10}},
\end{equation*}
where $T_* = 68~\rm mK$ and $A_{10} = 2.85 \times 10^{-15}\,\rm Hz$ represent the equivalent temperature and the Einstein coefficient for spontaneous emission in the hyperfine state, respectively. $n_i$ represents the number density of the species ``$i$'' present in the intergalactic medium (IGM) while $k^{i\rm H}_{10}$ represents
the corresponding collisional spin deexcitation rate. The deexcitation rates $k^{HH}_{10}$ and $k^{eH}_{10}$ can be approximated in a functional form as follows \cite{2001A&A...371..698L, 2006ApJ...637L...1K, Pritchard:2011xb, PhysRevD.110.123506}

\begin{alignat}{2}
	k^{HH}_{10} & = 3.1 \times 10^{-17}\left(\frac{T_{\rm gas}}{\mathrm{K}}\right)^{0.357}
	\cdot e^{-32\mathrm{K} / T_{\rm gas}}, \\
	\log_{10}{k^{eH}_{10}} & = -15.607 + \frac{1}{2}\log_{10}\left(\frac{T_{\rm gas}}{\mathrm{K}}\right)\times \nonumber \\ &
	\qquad\qquad~~\exp\left\{-\dfrac{\left[\log_{10} \left(T_{\rm gas}/\mathrm{K}\right)\right]^{4.5}}{1800}\right\},
\end{alignat}
Here, the $k^{iH}_{10}$ terms are expressed in the units of $\rm m^3\,s^{-1}$, and are
approximated under the consideration that $T_{\rm gas}<10^4\,\rm K$---which is reasonable
given that $T_{\rm gas}>10^4~\rm K$ at $z<200$ will represent a large emission signal during
the dark ages. Furthermore, at redshifts $z\lesssim 40$, $T_s$ approaches the CMB
temperature as $x_c$ becomes $\ll 1$, which leads to $T_{21}\sim 0$. However, Lyman alpha
$(\mathrm {Ly}\alpha)$ radiation originating from astrophysical structure formation during the
cosmic dawn can alter the spin temperature, as shown in Eq.~\eqref{eq:Ts}. Below, we
discuss the effect of star formation on $T_s$ and how it affects the $T_{21}$ signal.

\subsection{The cosmic dawn signal}
The onset of star formation during the cosmic dawn era $(z\lesssim 40)$ produces ionising radiation such as X-rays and Ly$\alpha$ that could heat, excite, and ionise the IGM. In the presence of Ly$\alpha$ radiation, $T_s$ recoupled to $T_{\rm gas}$ via the Wouthuysen-Field (WF) effect \cite{1952AJ.....57R..31W,1959ApJ...129..536F}, while the X-ray radiations heat and ionise the IGM. The Ly$\alpha$ coupling can be expressed as \cite{Hirata:2005mz, Mesinger:2010ne, Pritchard:2011xb},

\begin{equation}
    x_{\alpha} = \frac{T_{*}}{T_\gamma}\frac{4P_{\alpha}}{27 A_{10}},
\end{equation}
where $P_{\alpha}$ represents the total rate of Ly$\alpha$ photon scattering
per hydrogen atom. Furthermore, $P_{\alpha}$ depends on specific intensity $(J_{\alpha})$ of the Ly$\alpha$ photons, which can be calculated using the Ly$\alpha$ emissivity $\epsilon_{\alpha}$ \cite{Barkana:2000fd, Barkana:2004vb}. We first consider a fiducial Pop II star formation with a spectral energy distribution as $\phi(\alpha) = 2902.91 \tilde{E}^{-0.86}$ expressed in the units of $[\rm eV]^{-1}$ \cite{Mittal:2020kjs}. The term $\tilde{E} = E/E_{\rm ion}$ represents the photon energy relative to the ionisation energy $(E_{\rm ion} = 13.6\,\rm eV)$, where $E\in [E_{\alpha}, E_{\beta}]$, while $E_{\alpha} = 10.2\,\rm eV$ and $E_{\beta} = 12.09\,\rm eV$ represent the Ly$\alpha$ photon and Ly$\beta$ photon, respectively. We can now define $\epsilon_{\alpha}$ as \cite{Barkana:2000fd}

\begin{equation}
    \epsilon_{\alpha}(E, z) = f_{\alpha}\phi_{\alpha}(E) \frac{\dot{\rho}_{*}(z)}{m_b},
\end{equation}
where $m_b$, $f_{\alpha}$, and $\dot{\rho}_{*}$ represent the baryon mass, scaling parameter for $\phi_{\alpha}$, and star formation rate density (SFRD), respectively. The rate at which baryons collapse into dark matter haloes defines the SFRD \cite{Barkana:2004vb}.

We adopted the Press-Schechter formalism to determine the number of halo formations at redshift $z$ \cite{Press:1973iz}, which can be expressed as

\begin{equation}
    \dot{\rho}_{*}(z) = -f_{*} \bar{\rho}_{b}^{0}(1 + z)H(z) \frac{dF_{\rm coll}(z)}{dz},
\end{equation}
where $\bar{\rho}_{b}^{0} = \rho_c\Omega_{b,0}$ and $\rho_c$ represent baryon and critical density today, respectively. The term $f_{*}$ represents star formation efficiency. The fraction of baryons that have collapsed into dark matter haloes $(F_{\rm coll})$ is given by \cite{Barkana:2000fd}

\begin{equation}
    F_{\rm coll}(z) = \mathrm {erfc} \left[\frac{\delta_c(z)}{\sqrt{2}\sigma(m_{\rm min})}\right],
    \label{eq:F_coll}
\end{equation}
where $\delta_c$ is the linear critical density for collapse, $\sigma^2$ is the variance in the smoothed density field, and $\rm erfc[\cdot]$ represents complementary error function. The minimum halo mass $(m_{\rm min})$ that could host a star formation depends on virial temperature $(T_{\rm vir})$, which is expressed as \cite{Barkana:2000fd, Mohapatra:2023ury, Mohapatra:2025qpz}

\begin{equation}
	m_{\rm min} = \frac{10^8 \mathrm {M}_{\odot}}{\sqrt{\Omega_mh^2}}\left[\frac{10}{1+z}\frac{0.6}{\mu} \frac{\rm min(\rm T_{\rm vir})}{1.98\times 10^4}\right]^{3/2}\, ,
\end{equation}
where $\rm M_{\odot}$ represents the solar mass and $\mu\approx 1.22$ \cite{DAYAL20181}. The term $\rm min(T_{\rm vir}) = 10^4\, \rm K$ represents the minimum virial temperature of dark matter haloes hosting star formation. We used the \texttt{COLOSSUS} software to calculate $\delta_{c}/\sigma(m_{\rm min})$ \cite{Diemer:2017bwl}. After defining SFRD, we can now evaluate the Ly$\alpha$ specific intensity as \cite{Hirata:2005mz, Pritchard:2011xb},

\begin{equation}
    J_{\alpha} = \frac{c}{4\pi}(1+z)^2\sum_{n = 2}^{23}P_n \int_{z}^{z_{\rm max}}\frac{\epsilon_{\alpha}(E_n',z')}{H(z')}\, dz,
    \label{eq:J_alpha}
\end{equation}
where $P_n$ represents the probability at which an upper Lyman series photon redshifts to Ly$\alpha$ wavelength before getting absorbed or scattered. The tabulated values of $P_n$ can be found in articles \cite{Hirata:2005mz, Pritchard:2005an}. The redshifting energy of a photon $(E_n)$ originating at redshift $z$ will have $E_n'$ energy at redshift $z'$--- which can be expressed as $E_n' = E_n(1+z')/(1+z)$, here, $E_n$ represents the photon's transiting energy from $n^{\rm th}$ to the ground state of a HI atom. Thus, the upper limit of the integral in Eq. \eqref{eq:J_alpha} can be evaluated as \cite{Mittal:2020kjs}

\begin{equation}
    1+z_{\rm max} = \frac{E_{n+1}}{E_n}(1+z) = \frac{1-(1+n)^{-2}}{1- n^{-2}}(1+z).
\end{equation}

Finally, we rewrite the Ly$\alpha$ coupling coefficient as $x_{\alpha} = \frac{SJ_{\alpha}}{J_0}$, where $J_0 \approx 5.54\times 10^{-8}~\rm m^{-2}s^{-1}Hz^{-1}sr^{-1}$, and $S$ is called scattering correction, that we considered to be unity for this work. In the next section, we discuss the evolution of gas temperature in the presence of dynamical and X-ray heatings.

\section{Thermal evolution of IGM}\label{sec:Therma_evolution}

The thermal evolution of the universe in the absence of any exotic energy source can be expressed as \cite{Peebles:1968ja, Seager:1999bc, Seager:1999km, DAmico:2018sxd, Mitridate:2018iag, Short:2019twc}

\begin{alignat}{2}\label{eq:Gas_Evolution}
    \frac{dT_{\rm gas}}{dz} = 2\frac{T_{\rm gas}}{1+z} & + \frac{\Gamma_c}{(1+z)H(z)} (T_{\rm gas} - T_\gamma)\,\\ \nonumber &
     + \frac{2~Q_{\rm X}}{3n_bk_B(1+z)H(z)}\, ,
\end{alignat}
where $k_B$ is the Boltzmann constant and $H$ is the Hubble parameter. $n_{b} = n_H(1+f_{He}+x_e)$ represents the total number density of gas, where $f_{He} = 0.08$ and $x_e$ represent the helium fraction and ionization fraction, respectively \cite{PhysRevD.110.123506, Mohapatra:2023ury}. The first and second terms in the above equation account for the adiabatic cooling of the gas and inverse Compton scattering between CMB photons and free electrons present in the IGM. The third term contributes to the heating of the gas from the X-ray photons $(Q_{\rm X})$. The Compton scattering rate $(\Gamma_c)$ is expressed by

\begin{equation*}  
    \Gamma_c = \frac{8 x_e\sigma_T a_rT_{\gamma}^4 (z)}{3m_e c(1+f_{He}+x_e)}\, ,
    \label{Compton_scattering}
\end{equation*}
where $m_e$, $c$, and $\sigma_T$ represent the rest mass of an electron, the speed of light in vacuum, and Thomson scattering cross-section, respectively. Here, $a_r = 7.57\times 10^{-16}$~J\,$\text{m}^{-3}\,\text{K}^{-4}$ represents the radiation density constant. The ionisation evolution of the IGM in the absence of X-ray heating can be expressed as \cite{Peebles:1968ja, Seager:1999bc, PhysRevD.110.123506, Mohapatra:2023ury, Mohapatra:2025qpz}

\begin{equation}
	\frac{dx_e}{dz} = \frac{\mathcal{P}}{(1+z)H} \left[n_Hx_e^2\alpha_B - (1-x_e)\beta_Be^{-E_{\alpha}/k_BT_{\gamma}}\right],
	\label{xe_evolution}
\end{equation}

where $\mathcal{P}$ represents Peebles coefficient, while $\alpha_B$ and $\beta_B$ are the case-B recombination and photo-ionisation rates, respectively \cite{Seager:1999bc, Seager:1999km, Mitridate:2018iag}. The Peebles coefficient is given by \cite{Peebles:1968ja, DAmico:2018sxd}
\begin{equation*}
    \mathcal{P} = \frac{1+ \mathcal{K}_H\Lambda_Hn_H(1-x_e)}{1+ \mathcal{K}_H(\Lambda_H+\beta_H)n_H(1-x_e)},
\label{peeble_coefficient}
\end{equation*}
where $\mathcal{K}_H = \pi^2/(E_{\alpha}^3H)$, $E_\alpha = 10.2\,\rm eV$, and $\Lambda_H = 8.22\,\rm{sec}^{-1}$ represent redshifting Ly${\alpha}$ photons, rest frame energy of Ly$\alpha$ photon, and 2S-1S level two-photon decay rate in hydrogen atom respectively \cite{PhysRevA.30.1175}.

Now, let us discuss the role of X-ray photons in the evolution of the gas temperature and ionisation fraction. X-ray photons have a much longer mean free path than Ly$\alpha$ photons, allowing them to propagate far from their sources and effectively heat and partially ionise the gas \cite{Mirabel:2011rx}. The primary astrophysical sources of X-ray photons include X-ray binaries and mini-quasars \cite{Madau:2003um, Power:2012hm, Fragos:2013bfa}. The mechanism of X-ray heating proceeds as follows: X-ray photons traverse the gas and photoionize neutral hydrogen atoms, thereby releasing energetic free electrons. These electrons then transfer energy through excitations and collisions with other atoms and residual free electrons, thereby increasing the average kinetic energy of the gas and raising its temperature. To relate the X-ray emissivity with the SFR, we assume that the SFR is proportional to the rate at which baryonic matter collapses into virialised haloes, i.e., $dF_{\rm coll}/dt$ (see Eq.~\ref{eq:F_coll}). Following Ref.~\cite{Furlanetto:2006jb}, we express the X-ray heating term in Eq.~\eqref{eq:Gas_Evolution} as

\begin{equation}
    \frac{2}{3}\frac{Q_{\rm X}}{k_Bn_b(1+z)H(z)} = 5\times 10^5\,\mathrm{K}\,(f_Xf_*f_{Xh}) \frac{dF_{\rm coll}}{dz},
    \label{X-ray_term_in_Tg}
\end{equation}

where $f_X$ is a normalisation parameter (analogous to $f_{\alpha}$ for Ly$\alpha$ coupling), and $f_{Xh}$ is the fraction of X-ray energy deposited into heating the gas. As $f_X$ and $f_{Xh}$ are degenerate parameters, we treat their product $f_X f_{Xh}$ as a single effective quantity. Next, we investigate the effect of X-ray radiation on the ionisation fraction evolution. Since ionising photons are generated within galaxies, their production rate is considered to scale with the star formation rate \cite{Furlanetto:2006jb}. The ionisation efficiency $\xi_{\rm ion}$ can be expressed as

\begin{equation*}
    \xi_{\rm ion} = \mathrm{A_{He}}\, f_* f_{\rm esc} N_{\rm ion},
\end{equation*}

where $f_{\rm esc}$ denotes the fraction of ionising photons that escape their host galaxies, $N_{\rm ion}$ is the number of ionising photons produced per baryon, and $\mathrm{A_{He}} = 4/(4 - 3Y_p)$ accounts for the helium mass fraction. The evolution of the ionisation fraction in the presence of X-rays is given by \cite{Furlanetto:2006jb, Mohapatra:2023ury}

\begin{equation}
    \frac{dx_e}{dz} = \frac{dx_e}{dz}\bigg|_{\text{Eq.~\eqref{xe_evolution}}} + \xi_{\rm ion} \frac{dF_{\rm coll}}{dz}.
    \label{eq:xe_evolution_modified}
\end{equation}

Since $f_{\rm esc}$ and $N_{\rm ion}$ are degenerate, we treat their product $(f_{\rm esc} N_{\rm ion})$ as a single parameter and set it to unity for simplicity.

In the presence of dynamical heating due to the drag between MACHOs and the wake, the temperature of the IGM will increase during the post-recombination epoch. This extra energy dissipation into the IGM can be expressed as 

\begin{equation}
    \frac{dT_{\rm gas}}{dz} = \frac{dT_{\rm gas}}{dz}\Bigg{|}_{\text{Eq.}~\eqref{eq:Gas_Evolution}} - \frac{2}{3(1+z)H(z)n_b}\frac{dE}{dVdt}~,
    \label{eq:thermal_IGM_evolution}
\end{equation}
where the volumetric energy dissipation rate term $dE/dVdt$ is taken from Eq. \eqref{eq:energy_dissipation}. On the other hand, the ionisation evolution of the IGM follows Eq. \eqref{eq:xe_evolution_modified}, as the energy dissipation does not ionise the IGM directly.

\section{Results}\label{sec:Result}

\begin{figure*}
    \begin{center}
        \subfloat[]{\includegraphics[width = 0.45\textwidth]{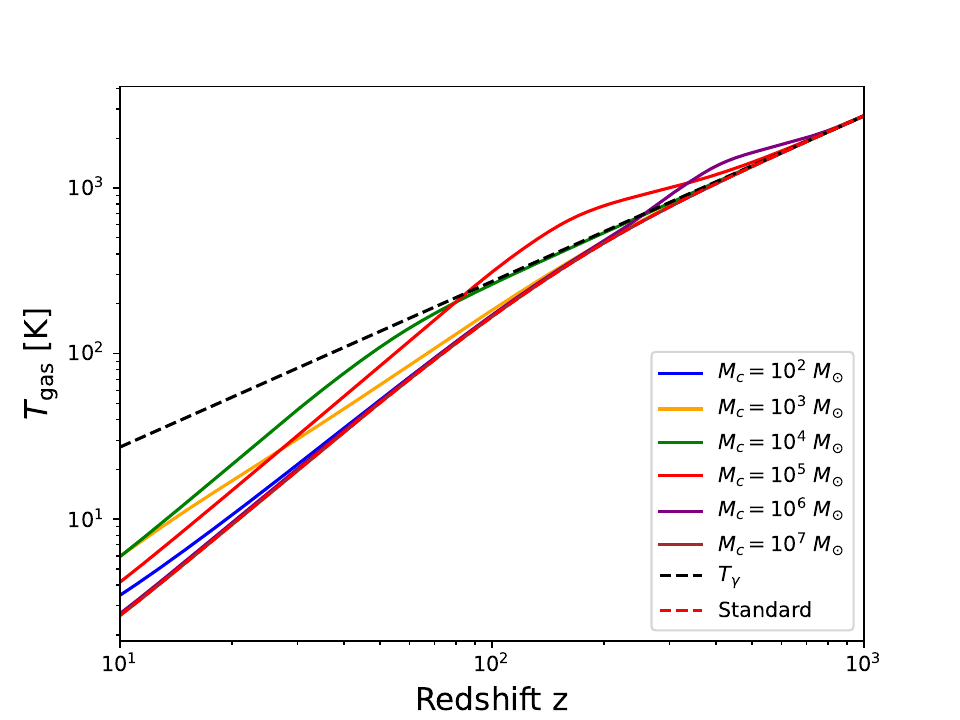}\label{fig:IGM_evolution}}
        \subfloat[]{\includegraphics[width = 0.45\textwidth]{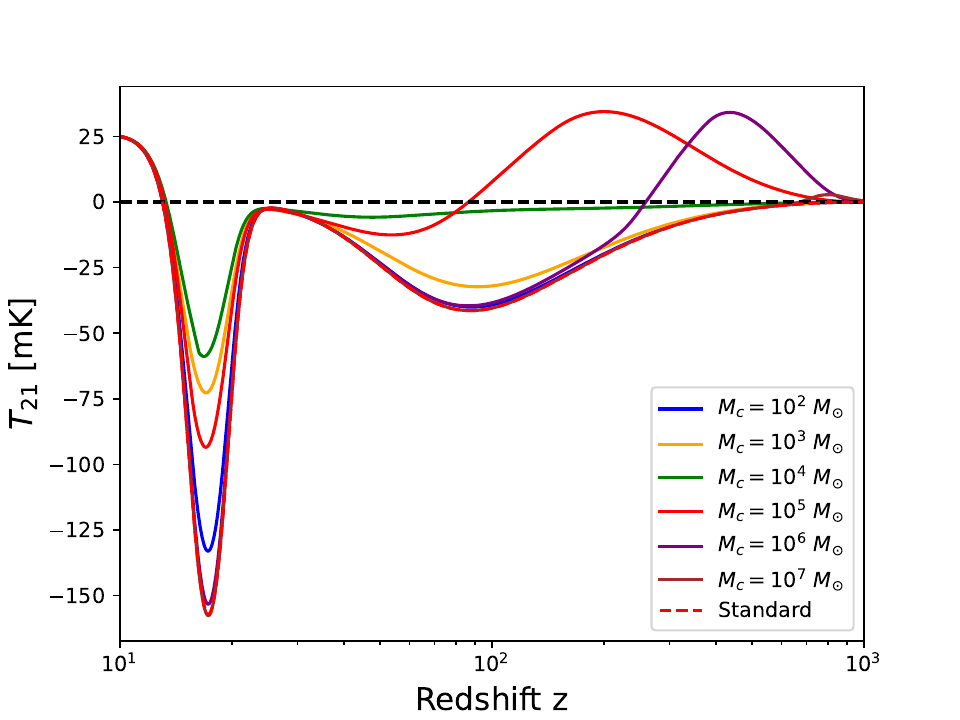}\label{fig:21-cm_evolution}}
        \vfill
        \subfloat[]{\includegraphics[width = 0.45\textwidth]{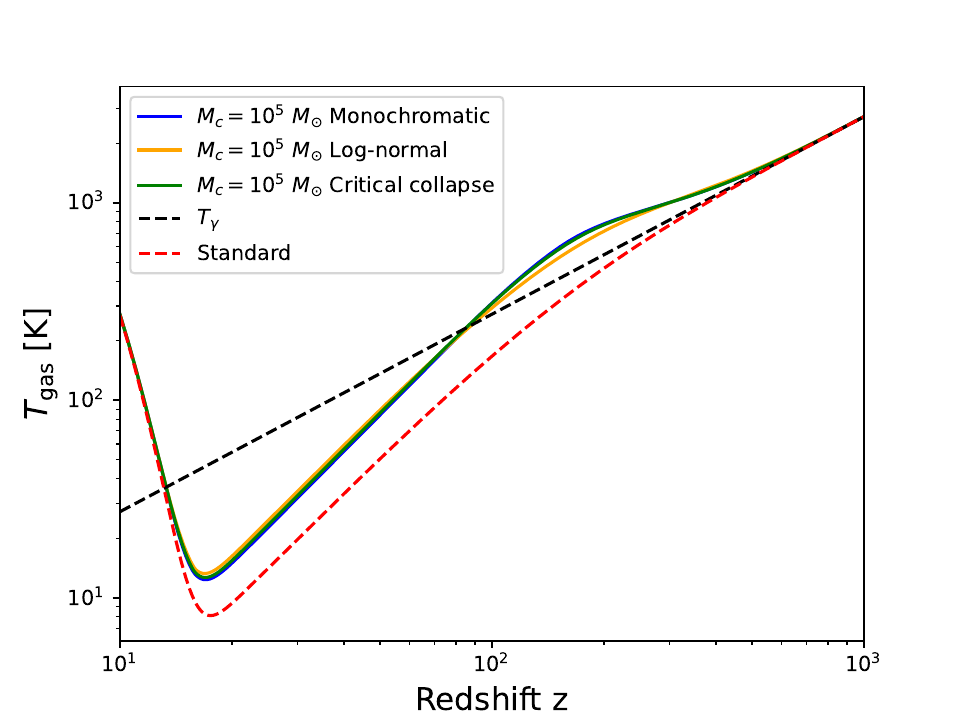}\label{fig:IGM_evolution_extended}}
        \subfloat[]{\includegraphics[width = 0.5\textwidth]{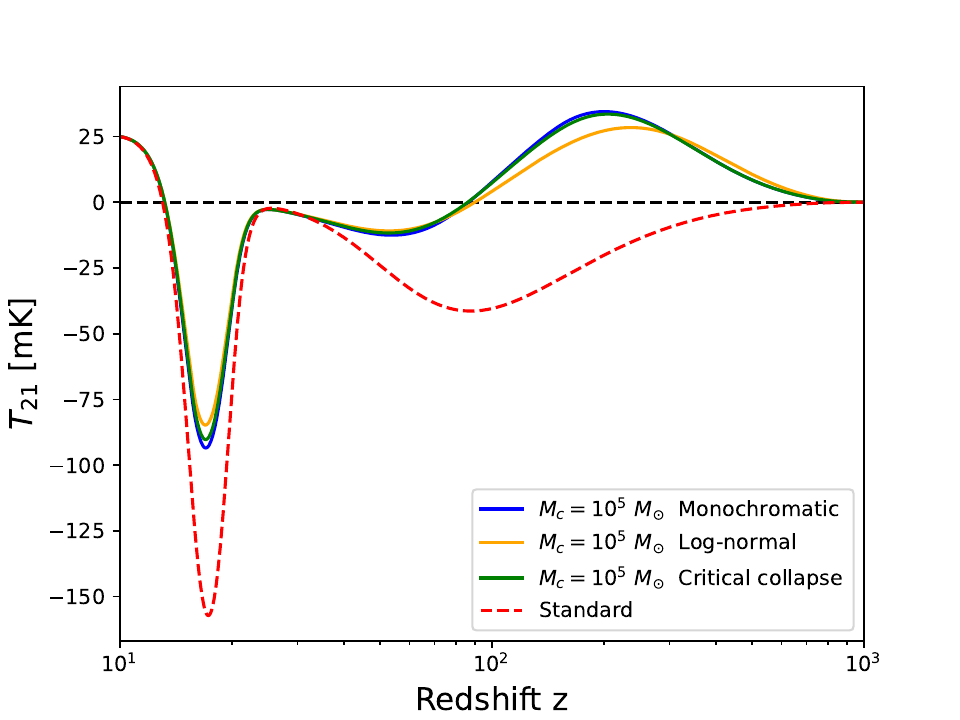}\label{fig:21-cm_evolution_extended}}
    \end{center}
    
\caption{\emph{Top panel:} Shows the effect of dynamical heating on the thermal evolution of the IGM and 21-cm signal, for monochromatic distributions with masses ranging $(10^2-10^7)~M_\odot$. The black and red dashed lines on the left-side figure represent $T_\gamma(z)$ and $T_{\rm gas} (z)$ in a $\Lambda$CDM framework. The effects of dynamical heating for different masses with $f_M = 1$ in the absence of X-ray heating are colour-coded in the legend, whereas the right-side figure presents the corresponding $T_{21}$ signals. \emph{Bottom panel:} Shows the comparison between the three distribution functions--- monochromatic, log-normal with $\sigma = 1$, and critical collapse--- for a fixed $M_c = 10^5~M_\odot$.}
\label{fig:TM_T21_extended}
\end{figure*}

To evaluate the thermal evolution of IGM in the absence of dynamical heating, we simultaneously solve Eq. \eqref{eq:Gas_Evolution} and \eqref{eq:xe_evolution_modified} with the initial conditions $T_{\rm gas} = 2892~\rm K$ and $x_e = 0.0976$ at the drag redshift $z = 1060$, taken from Recfast++ \cite{10.1111/j.1365-2966.2010.16940.x, 10.1111/j.1365-2966.2010.17940.x}. In Fig. \ref{fig:IGM_evolution}, the black and red dashed lines represent $T_{\gamma}(z)$ and $T_{\rm gas}(z)$ without X-ray heating in a $\Lambda$CDM framework, with $T_{\rm gas}$ equal to $\sim 6.8~\rm K$ and $\sim 138.83~\rm K$ at $z = 17$ and $89$, respectively. 

Further, to include energy dissipation from dynamical heating, we solve Eq. \eqref{eq:velocity_evolution} with initial condition $v_{bc} = 30~\rm kms^{-1}$ at $z\sim 1060$ along with equations \eqref{eq:xe_evolution_modified} and \eqref{eq:thermal_IGM_evolution}. We first consider a monochromatic distribution of MACHOs with masses varying from $10^2$ to $10^7~M_{\odot}$ with $f_M = 1$ --- as shown in Fig. \ref{fig:IGM_evolution}. From Eq. \eqref{eq:dynamical_force} we can observe that $F_{\rm DF}\propto M^2/v_{bc}^2$, thus a smaller mass generates less drag compared to larger masses. Furthermore, from Eq. \eqref{eq:velocity_evolution} we observe that, $v_{bc}$ evolution is proportional to $F_{\rm DF}/MH(z)$ which translates the proportionality to $M~(1+z)^{-5/2}$ in a matter-dominated era, along with a $(1+z)$ fall due to Hubble expansion. Therefore, a larger mass can decelerate faster compared to a smaller mass, reducing $v_{bc}$ faster and entering a subsonic regime from a supersonic one. In terms of energy budget, the MACHOs have maximum kinetic energy at the drag epoch, and as the $v_{bc}$ falls, this energy is dissipated as heat into the IGM. Here, we can also analyse that a smaller mass will alter the IGM temperature later in time compared to the heavier masses, as an integrated effect--- as shown in Eq. \eqref{eq:energy_dissipation}. In Fig. \ref{fig:IGM_evolution}, the blue and purple solid lines depict the same, where the rate of energy dissipation shifts from lower to higher redshift as we increase the mass. Note that the masses heavier than $10^7~M_\odot$ should alter $T_{\rm gas}$ at redshifts $z\gtrsim z_d$ without affecting the thermal evolution at lower redshifts. However, at such redshift these objects have a $\mathcal{M} \sim \sqrt{3}\times 10^{-5}$ before $z_d$, placing them in a subsonic regime, thus making the energy dissipation nearly negligible.

We then evaluate the evolution of the global 21-cm signal in the presence of dynamical heating. First, we enabled X-ray heating and Ly$\alpha$ coupling to obtain an absorption amplitude during the cosmic dawn era $(z\lesssim 20)$. As the cosmic dawn era is sensitive to star formation history, the absorption amplitude of the global 21-cm signal depends on many astrophysical parameters. In this work, we fixed the $\rm T_{ vir}$, $f_X$, $f_*$, and $f_\alpha$ to $10^4~\text{K}$, $1.22$, $0.1$, and $5$, respectively, to obtain $T_{21}\sim -153~\rm mK$ at $z\sim 17$. In Fig. \ref{fig:21-cm_evolution}, the red dashed line represents the $T_{21}$ signal in a $\Lambda$CDM framework, whereas the brown solid line represents heating from MACHOs with mass $10^7~M_\odot$. We can observe that dynamical energy from such objects is dissipated at early redshifts $(z\gtrsim 600)$, producing an emission signal and with almost no effect on later redshifts. By decreasing the mass to $10^6$ and $10^5~M_\odot$ depicted in solid purple and red lines, respectively, we observe a significant emission signal during the dark ages $(z\gtrsim 30)$. Particularly, for a mass of $10^4~M_\odot$, the dark ages signal vanishes, leaving an absorption signal during the cosmic dawn era. On further decreasing the mass, the dynamical energy is dissipated at later redshifts, leaving a mild effect on the dark ages while reducing the cosmic dawn signal. Before we specify the $T_{21}$ thresholds and constrain the heating, let us discuss the evolution of $T_{\rm gas}$ and $T_{21}$ for extended mass distribution.

In this work, we consider log-normal and critical collapse distributions. To evaluate the thermal evolution of IGM and $T_{21}$, we first define the velocity evolution [see Eq. \eqref{eq:velocity_evolution}] for each mass bin as $v_{bc}(M')$. We then set $f_M =1$ and solve Eqs. \eqref{eq:xe_evolution_modified} and \eqref{eq:thermal_IGM_evolution}, along with the entire $v_{bc}(M')$ bins simulataneously. Note that we assume the initial condition for all $v_{bc}(M')$ is set to the rms value of $v_{bc} = 30~\rm kms^{-1}$, suggesting that all MACHOs have the same relative velocity with respect to the baryon during the drag epoch. Specifically, we consider $v_{bc}(M')$ for $50$ mass bins to track the energy dissipation from each mass. In both extended mass distributions, the mass term $(M)$ becomes an integrating variable, while $M_c$ regulates the distribution function. Therefore, from here onwards, we define the MACHOs' mass as $M_c$ instead of $M$ to illustrate the effects of dynamical heating. In Fig. \ref{fig:IGM_evolution_extended}, the red dashed line represents $T_{\rm gas}$ in the presence of X-ray heating, and the solid orange and green lines represent IGM evolution for log-normal with $\sigma =1$ and critical collapse distributions, respectively, for $M_c = 10^5~M_\odot$. For comparison, we present the IGM evolution in the presence of MACHOs with mass $M_c = 10^5~M_\odot$ in a monochromatic distribution--- depicted in blue solid line. The presence of energy dissipation from MACHOs can heat the IGM above CMB temperature during the dark ages. We can observe that the $T_{\rm gas}$ for the monochromatic and critical collapse distribution nearly overlap. This can be interpreted by analysing the mass distribution for critical collapse [see Eq. \eqref{eq:critical_collapse}] which suggests that the number density of MACHOs with $M\neq M_c$ is heavily suppressed, thus making MACHOs with mass $M_c$ the primary source of energy dissipation. On the other hand, the sudden rise in $T_{\rm gas}$ at $z\lesssim 20$ is due to the additional X-ray heating from star formation during the cosmic dawn era. Lastly, we present the corresponding $T_{21}$ signals in Fig. \ref{fig:21-cm_evolution_extended}, where the red dashed line represents the $T_{21}$ signal in a $\Lambda$CDM framework in the presence of the aforementioned star formation scenario. We can clearly observe that the effect of heating from the critical collapse and monochromatic distributions overlap, while the log-normal heats the IGM comparatively less.

\begin{figure}
    \centering
    \includegraphics[width=\linewidth]{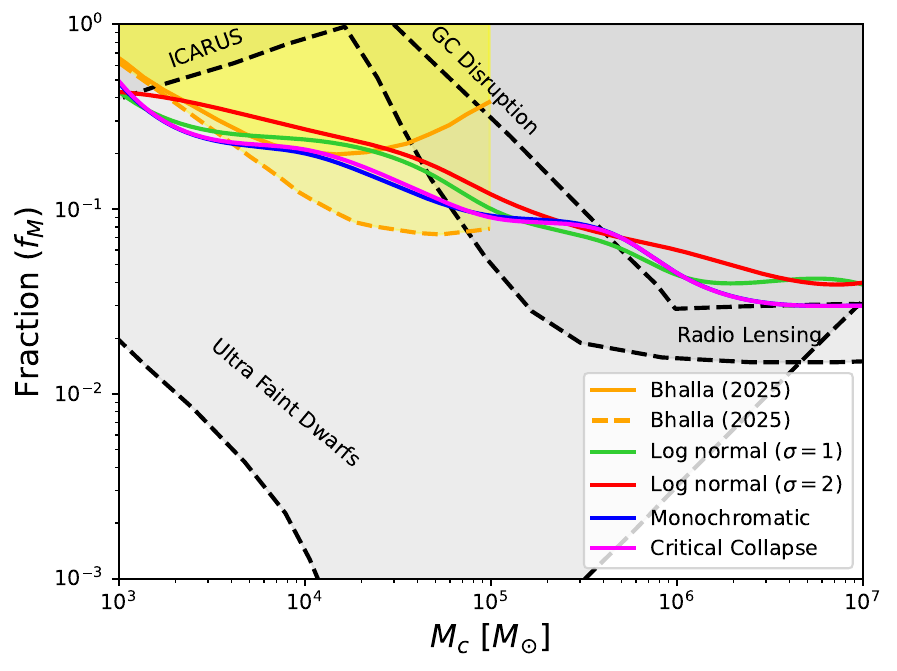}
    \caption{Presents the upper bounds on the fraction of dark matter residing inside MACHOs with respect to their masses. The blue and magenta solid lines represent a monochromatic and critical collapse distribution, respectively. The solid green and red lines represent log-normal distributions with $\sigma=1$ and $2$, respectively. The yellow solid and dashed lines are derived by limiting the cosmic dawn global 21-cm signal and the power spectrum \cite{Bhalla:2025orw}. The black dashed lines are the existing bounds from non-detection of lensing effects from compact radio sources \cite{Zhou:2021tvp}, from the nondisruption of galaxy clusters \cite{1993ApJ...413L..93M}, the caustic crossing of ICARUS \cite{Oguri:2017ock}, and from the observed half-light radius of UFDs \cite{Graham:2023unf}.}
    \label{fig:bounds}
\end{figure}
Before presenting the bounds on the fraction of dark matter residing inside MACHOs with respect to their masses using the $T_{21}$ signal, we define the exclusion criteria. To restrict the dynamical heating we consider all of the following thresholds to be simultaneously true: $\Delta T_{21}\equiv T_{21}^{\rm MACHO}~-T_{21}^{ \Lambda\rm CDM}$ to be $50~\rm mK$ and $15~\rm mK$ at $z\sim 17$ and $\sim 89$, respectively, and no emission signal at $z\gtrsim 300$. We consider these specific redshifts for the following reasons: (1) EDGES tentative observation suggest an absorption feature centered at $z\sim 17.2$ \cite{Bowman:2018yin}, (2) the authors in Ref. \cite{Bevins:2025gjv} suggest the dark ages absorption amplitude peaks at $z\sim 85$ after simultaneously analysing Planck and the Wilkinson Microwave Anisotropy Probe (WMAP), and two late time probes, Dark Energy Survey galaxy lensing and clustering and baryon acoustic oscillations, and (3) as the $\Lambda\rm CDM$ framework suggests a strong thermal coupling between baryon and CMB at $z\gtrsim 300$, we restricted the fraction of dark matter to produce any emission signal.

In Fig. \ref{fig:bounds}, the blue solid lines represent the upper bound on $f_M$ for different masses for a monochromatic distribution. We observe that for $M_c = 10^3~M_\odot$ and $10^5~M_\odot$, $f_M$ attains $\sim 0.48$ and $0.13$, respectively; however, the former is constrained by the cosmic dawn signal, while the dark ages constrain the latter. On the other hand, $M_c =  10^7~M_\odot$ lowers the $f_M\sim 0.03$ to restrict $T_{21}\lesssim0~\rm mK$ at $z\gtrsim 300$. Thus, as we shift from lower to higher masses, the sensitivity of the $f_M$ bound shifts from the cosmic dawn to the higher redshift emission signal via the dark ages signal. Moreover, the heavier MACHOs tend to decelerate faster compared to smaller ones, thereby dissipating the dynamical energy earlier in time. We then show the bounds from log-normal distributions with $\sigma = 1$ and $2$ in green and red solid lines, respectively. Lastly, the magenta solid line represents bounds from the critical collapse distribution. On comparing the three distribution functions, we find that the bounds on $f_M$ for a log-normal distribution become mildly stronger as we reduce the variance from $\sigma = 2$ to $1$--- as shown in the red and green solid lines, respectively. This can be interpreted by analysing the fact that reducing the variance reduces the number densities of MACHOs near the characteristic mass $M_c$ significantly, thereby governing the dynamical energy dissipation into the IGM primarily on $M_c$. Therefore, the monochromatic and critical collapse distributions yield a similar bound and a stronger bound on $f_M$--- as shown in the blue and magenta lines, respectively.

To further compare our results with the existing astrophysical bounds, we present the following findings in the literature: 
\begin{enumerate}
    
    \item In Ref. \cite{Oguri:2017ock}, the authors use the discovery of MACS J1149 Lensed Star 1 (LS1), also known as ``ICARUS'', as the primary observational evidence to constrain the fraction of MACHOs. Here, they argued the non-detection of possible microcaustics in the presence of compact dark matter objects suggests that dark matter in the mass range $10^{-5}\lesssim M/M_\odot\lesssim 10^2$ is smooth or non-clustered in nature.

    \item In Ref. \cite{1993ApJ...413L..93M}, the authors establish an upper bound on the mass of compact dark matter objects based on the survival of globular clusters (GCs) in the Milky Way halo. Massive compact objects transfer energy to stars through gravitational interaction, increasing the stellar velocity and driving the cluster toward disruption. Here, they argued that if compact objects like black holes constitute 100\% of dark matter, then their mass greater than $10^3~M_\odot$ would heat several low-mass halo globular clusters to the point of disruption within half of their lifetime--- thus constraining the fraction of dark matter as compact objects.

    \item The author in Ref. \cite{Zhou:2021tvp} constrain the abundance of supermassive compact objects through the non-detection of milli-lensing events in compact radio sources. Considering such objects as possible progenitors of the supermassive black holes observed at redshifts $ z \gtrsim 6$, the absence of any confirmed lensing signatures places constraints on compact objects in the mass range $(10^4-10^8)~M_\odot$.

    \item Authors in Ref. \cite{Graham:2023unf} used the observed half-light radius of stars in the Segue-I ultrafaint dwarf (UFD) galaxy. A half-light radius refers to the radius within which half of an object's total luminosity is emitted. Here, the authors consider gravitational interaction between the MACHOs and stars, where they have considered the gas within MACHOs to be typically hotter than the stars. This allows MACHOs to transfer energy, resulting in expansion of the stellar gas within the stars and hence affecting the half-light radius.    
\end{enumerate}

Lastly, in the context of cosmologically derived bounds, we present upper bounds for monochromatic distribution derived using the cosmic dawn global 21-cm signal and its power spectrum by the authors in Ref. \cite{Bhalla:2025orw}. The authors considered $T_{21}^{\rm MACHO}>T_{21}^{\Lambda\rm CDM}+50~\rm mK$ and dimensionless power spectrum $\Delta^2_{T_{21}}>92.7~\rm mK^2$ at redshift $z\sim 17$ depicted in solid and dashed yellow lines, respectively. We note that our derived bounds are tighter than those derived by the authors with the same threshold condition, that is, $T_{21}^{\rm MACHO}>T_{21}^{\Lambda\rm CDM}+50~\rm mK$. On a closer inspection, we find that the upper bounds derived in this work (shown in the blue solid line) are symmetrically shifted to a lower $f_M$ value with a difference of $0.1$ for $M_c\lesssim 10^4~M_\odot$ from the bounds derived by the authors (shown in the yellow solid line). This is because bounds on such masses are sensitive to the absorption amplitude of $T_{21}$ during the cosmic dawn, which depends on star formation history. However, as the larger masses tend to affect the dark ages signal, the bounds derived in this work diverge from the previous work and become stronger. Additionally, the inclusion of dark ages constraints provides a relatively cleaner window than the cosmic dawn on a wider range of masses.

\section{Conclusion and discussion}\label{sec:conclusion}
We have studied the impact of dynamical friction heating from massive compact halo objects on the thermal and ionisation evolution of IGM and its imprint on the global 21-cm signal, during both the dark ages and the cosmic dawn. MACHOs stream through baryonic fluid during the post-drag epoch at a bulk relative velocity of $v_{bc} \sim 30\,\text{km\,s}^{-1}$ (the rms value at $z \sim 1060$), induce a drag force and dissipate energy into the IGM at a rate determined by Eq. \eqref{eq:energy_dissipation}. For a monochromatic mass spectrum, the heating is most efficient in the mass range $10^4 \lesssim M/M_\odot \lesssim 10^7$, which deposits energy predominantly during the dark ages $(30 \lesssim z \lesssim 600)$ and produces a prominent emission signal in $T_{21}$. Masses above $\sim 10^7\,M_\odot$ decelerate rapidly into the subsonic regime before recombination, where the drag force is strongly suppressed, and leave almost no imprint at $z\lesssim 1100$. Masses below $\sim 10^3\,M_\odot$ dissipate their energy at late redshifts and mildly reduce the cosmic dawn absorption amplitude.

Applying the exclusion criteria defined in Sec. \eqref{sec:Result}, we obtain upper bounds on the MACHO dark matter fraction as a function of characteristic mass, shown in Fig. \ref{fig:bounds}. For a monochromatic distribution, the strongest bound $f_M \lesssim 0.03$ is obtained at $M_c = 10^7\,M_\odot$, where demanding no emission signal ($T_{\rm gas}\lesssim T_{\gamma}$) at $z \gtrsim 300$. At the intermediate masses $(M_c \sim 10^5\,M_\odot)$, the constraint weakens and is instead set by the dark ages absorption amplitude at $z \sim 89$. At lower masses $(M_c \lesssim 10^4\,M_\odot)$, the bounds on $f_M$ become sensitive to the cosmic dawn era. We then considered extended mass distributions, and for a fixed characteristic mass $M_c$, the log-normal distribution produces less pronounced heating than the monochromatic and critical collapse case, because the number densities of MACHOs distributed near $M_c$ reduces significant that contribute little to no additional energy dissipation. This results in bounds shallower than the monochromatic case.

Our dark-ages-based constraints are stronger than the cosmic-dawn bounds derived in Ref. \cite{Bhalla:2025orw} under the same $\Delta T_{21} > 50\,\text{mK}$ threshold at $z \sim 17$, with a systematic shift of approximately 0.1 in $f_M$ at $M_c \lesssim 10^4\,M_\odot$. The improvement grows with mass $M_c \gtrsim 10^4\,M_\odot$, where the dark ages signal provides a substantially stronger and qualitatively different probe, as it remains insensitive to the astrophysical parameters governing star formation. Our bounds are stronger than the astrophysical constraints from microlensing (ICARUS), globular cluster disruption, and complementary to the milli-lensing of compact radio sources and the half-light radii of ultrafaint dwarf galaxies-based constraints.

Before concluding this section, we note that several simplifying assumptions considered in this work are worth further investigation. We adopt the rms value of $ v_{bc}$ as a spatially uniform initial condition, whereas the true velocity field follows a Boltzmann distribution with coherence on scales $\sim 3\,\text{Mpc}$; a proper treatment would account for the spatial variance in heating rate which has been partly addressed for a monochromatic distribution by authors in Ref. \cite{Bhalla:2025orw} for the cosmic dawn era. We have considered a speculative extended mass distribution assuming MACHOs to follow a PBH-like distribution, while a physically motivated scenario will be worth exploring. Both the X-ray heating prescription and the Ly$\alpha$ coupling depend on uncertain astrophysical parameters, and our cosmic-dawn bounds carry those astrophysical uncertainties. We have also neglected accretion effects, which may become relevant for the most massive objects considered. In future work, we will extend our analysis by including the accretion effect on the dynamical friction and also calculate the dark ages 21-cm power spectrum that may yield tighter constraints. With forthcoming observations from REACH, LuSee-Night, and other next-generation 21-cm experiments, particularly targeting the dark ages signal, the constraints derived here could be substantially tightened or provide a detection.

\section{Data availability}
The code underlying this work will be made publicly available on this \href{https://github.com/vivekanandmohapatra/Dynamical-Heating-from-MACHOs-and-21-cm-Signal/tree/main}{GitHub} repository in due course. In the meantime, the code can be obtained from the corresponding author upon reasonable request.

\section*{Acknowledgements}
V.M would like to thank Badal Bhalla and Sanjeev Kalita for fruitful discussions, suggestions, and clarifications. We thank the anonymous reviewers for the suggestions, corrections, and improving the presentation of the manuscript. We also acknowledge the use of \texttt{Numpy} \cite{harris2020array} and \texttt{Scipy} \cite{2020SciPy-NMeth} for the numerical evaluations, and \texttt{Matplotlib} \cite{Hunter:2007} for visualisations. A.C.N would like to acknowledge financial support from ISRO RAC-S, proposal no: RAC-S/GU/2024/4/56, India.

\bibliography{main}

@article{Bevins:2025gjv,
    author = "Bevins, Harry T. J.",
    title = "{The Magnitude of the Dark Ages 21 cm Signal in the Context of Existing Early and Late Time Constraints on {\ensuremath{\Lambda}}CDM}",
    eprint = "2512.11568",
    archivePrefix = "arXiv",
    primaryClass = "astro-ph.CO",
    doi = "10.3847/2041-8213/ae356b",
    journal = "Astrophys. J. Lett.",
    volume = "997",
    number = "2",
    pages = "L35",
    year = "2026"
}

@misc{Bhalla:2025orw,
      title={Dynamical Heating from Dark Compact Objects and Axion Minihalos: Implications for the 21-cm Signal}, 
      author={Badal Bhalla and Aurora Ireland and Hongwan Liu and Huangyu Xiao and Tao Xu},
      year={2025},
      eprint={2512.00169},
      archivePrefix={arXiv},
      primaryClass={astro-ph.CO},
      url={https://arxiv.org/abs/2512.00169}, 
}

@article{Ramirez:2022mys,
    author = "Ramirez, Edward D. and Buckley, Matthew R.",
    title = "{Constraining dark matter substructure with Gaia wide binaries}",
    eprint = "2209.08100",
    archivePrefix = "arXiv",
    primaryClass = "hep-ph",
    doi = "10.1093/mnras/stad2583",
    journal = "Mon. Not. Roy. Astron. Soc.",
    volume = "525",
    number = "4",
    pages = "5813--5830",
    year = "2023"
}

@article{Zhou:2021tvp,
    author = "Zhou, Huan and Lian, Yujie and Li, Zhengxiang and Cao, Shuo and Huang, Zhiqi",
    title = "{Constraints on the abundance of supermassive primordial black holes from lensing of compact radio sources}",
    eprint = "2106.11705",
    archivePrefix = "arXiv",
    primaryClass = "astro-ph.CO",
    doi = "10.1093/mnras/stac915",
    journal = "Mon. Not. Roy. Astron. Soc.",
    volume = "513",
    number = "3",
    pages = "3627--3633",
    year = "2022"
}

@article{Graham:2023unf,
    author = "Graham, Peter W. and Ramani, Harikrishnan",
    title = "{Constraints on dark matter from dynamical heating of stars in ultrafaint dwarfs. I. MACHOs and primordial black holes}",
    eprint = "2311.07654",
    archivePrefix = "arXiv",
    primaryClass = "hep-ph",
    doi = "10.1103/PhysRevD.110.075011",
    journal = "Phys. Rev. D",
    volume = "110",
    number = "7",
    pages = "075011",
    year = "2024"
}

@ARTICLE{1993ApJ...413L..93M,
       author = {{Moore}, Ben},
        title = "{An Upper Limit to the Mass of Black Holes in the Halo of the Galaxy}",
      journal = {Astrophys. J. Lett.},
         year = 1993,
        month = aug,
       volume = {413},
        pages = {L93},
          doi = {10.1086/186967},
archivePrefix = {arXiv},
       eprint = {astro-ph/9306004},
 primaryClass = {astro-ph},
       adsurl = {https://ui.adsabs.harvard.edu/abs/1993ApJ...413L..93M}
}

@article{Bowman:2018yin,
    author = "Bowman, Judd D. and Rogers, Alan E. E. and Monsalve, Raul A. and Mozdzen, Thomas J. and Mahesh, Nivedita",
    title = "{An absorption profile centred at 78 megahertz in the sky-averaged spectrum}",
    eprint = "1810.05912",
    archivePrefix = "arXiv",
    primaryClass = "astro-ph.CO",
    doi = "10.1038/nature25792",
    journal = "Nature",
    volume = "555",
    number = "7694",
    pages = "67--70",
    year = "2018"
}

@ARTICLE{2001A&A...371..698L,
       author = {{Liszt}, H.},
        title = "{The spin temperature of warm interstellar H I}",
      journal = "Astron. Astrophys.",
         year = 2001,
        month = may,
       volume = {371},
        pages = {698-707},
          doi = {10.1051/0004-6361:20010395},
archivePrefix = {arXiv},
       eprint = {astro-ph/0103246},
 primaryClass = {astro-ph}
}

@ARTICLE{2006ApJ...637L...1K,
       author = {{Kuhlen}, Michael and {Madau}, Piero and {Montgomery}, Ryan},
        title = "{The Spin Temperature and 21 cm Brightness of the Intergalactic Medium in the Pre-Reionization era}",
      journal = "Astrophys. J. Lett.",
         year = 2006,
        month = jan,
       volume = {637},
       number = {1},
        pages = {L1-L4},
          doi = {10.1086/500548},
archivePrefix = {arXiv},
       eprint = {astro-ph/0510814},
 primaryClass = {astro-ph}
}

@article{Mohapatra:2023ury,
    author = "Mohapatra, Vivekanand and Johnny, J. and Natwariya, Pravin Kumar and Goswami, Jishnu and Nayak, Alekha C.",
    title = "{In search of global 21-cm signal using artificial neural network in light of ARCADE 2}",
    eprint = "2306.02039",
    archivePrefix = "arXiv",
    primaryClass = "astro-ph.CO",
    doi = "10.1140/epjc/s10052-025-14256-0",
    journal = "Eur. Phys. J. C",
    volume = "85",
    number = "5",
    pages = "533",
    year = "2025"
}

@article{Madau:2003um,
    author = "Madau, Piero and Rees, Martin J. and Volonteri, Marta and Haardt, Francesco and Oh, S. Peng",
    title = "{Early reionization by miniquasars}",
    eprint = "astro-ph/0310223",
    archivePrefix = "arXiv",
    doi = "10.1086/381935",
    journal = "Astrophys. J.",
    volume = "604",
    pages = "484--494",
    year = "2004"
}

@article{Power:2012hm,
    author = "Power, Chris and James, Gillian F. and Combet, Celine and Wynn, Graham",
    title = "{Feedback from High-Mass X-Ray Binaries on the High Redshift Intergalactic Medium : Model Spectra}",
    eprint = "1211.5854",
    archivePrefix = "arXiv",
    primaryClass = "astro-ph.CO",
    doi = "10.1088/0004-637X/764/1/76",
    journal = "Astrophys. J.",
    volume = "764",
    pages = "76",
    year = "2013"
}

@article{Fragos:2013bfa,
    author = "Fragos, Tassos and Lehmer, Bret D. and Naoz, Smadar and Zezas, Andreas and Basu-Zych, Antara R.",
    title = "{Energy Feedback from X-ray Binaries in the Early Universe}",
    eprint = "1306.1405",
    archivePrefix = "arXiv",
    primaryClass = "astro-ph.CO",
    doi = "10.1088/2041-8205/776/2/L31",
    journal = "Astrophys. J. Lett.",
    volume = "776",
    pages = "L31",
    year = "2013"
}

@article{Pritchard:2005an,
    author = "Pritchard, Jonathan R. and Furlanetto, Steven R.",
    title = "{Descending from on high: lyman series cascades and spin-kinetic temperature coupling in the 21 cm line}",
    eprint = "astro-ph/0508381",
    archivePrefix = "arXiv",
    doi = "10.1111/j.1365-2966.2006.10028.x",
    journal = "Mon. Not. Roy. Astron. Soc.",
    volume = "367",
    pages = "1057--1066",
    year = "2006"
}

@article{Mirabel:2011rx,
    author = "Mirabel, I. F. and Dijkstra, M. and Laurent, P. and Loeb, A. and Pritchard, J. R.",
    title = "{Stellar black holes at the dawn of the universe}",
    eprint = "1102.1891",
    archivePrefix = "arXiv",
    primaryClass = "astro-ph.CO",
    doi = "10.1051/0004-6361/201016357",
    journal = "Astron. Astrophys.",
    volume = "528",
    pages = "A149",
    year = "2011"
}

@article{Mitridate:2018iag,
    author = "Mitridate, Andrea and Podo, Alessandro",
    title = "{Bounds on Dark Matter decay from 21 cm line}",
    eprint = "1803.11169",
    archivePrefix = "arXiv",
    primaryClass = "hep-ph",
    doi = "10.1088/1475-7516/2018/05/069",
    journal = "JCAP",
    volume = "05",
    pages = "069",
    year = "2018"
}

@article{Seager:1999bc,
    author = "Seager, Sara and Sasselov, Dimitar D. and Scott, Douglas",
    title = "{A new calculation of the recombination epoch}",
    eprint = "astro-ph/9909275",
    archivePrefix = "arXiv",
    doi = "10.1086/312250",
    journal = "Astrophys. J. Lett.",
    volume = "523",
    pages = "L1--L5",
    year = "1999"
}

@article{Seager:1999km,
    author = "Seager, Sara and Sasselov, Dimitar D. and Scott, Douglas",
    title = "{How exactly did the universe become neutral?}",
    eprint = "astro-ph/9912182",
    archivePrefix = "arXiv",
    reportNumber = "UBC-COS-99-07",
    doi = "10.1086/313388",
    journal = "Astrophys. J. Suppl.",
    volume = "128",
    pages = "407--430",
    year = "2000"
}

@article{DAmico:2018sxd,
    author = "D'Amico, Guido and Panci, Paolo and Strumia, Alessandro",
    title = "{Bounds on Dark Matter annihilations from 21 cm data}",
    eprint = "1803.03629",
    archivePrefix = "arXiv",
    primaryClass = "astro-ph.CO",
    reportNumber = "CERN-TH-2018-052, IFUP-TH-2018",
    doi = "10.1103/PhysRevLett.121.011103",
    journal = "Phys. Rev. Lett.",
    volume = "121",
    number = "1",
    pages = "011103",
    year = "2018"
}

@article{Peebles:1968ja,
    author = "Peebles, P. J. E.",
    title = "{Recombination of the Primeval Plasma}",
    doi = "10.1086/149628",
    journal = "Astrophys. J.",
    volume = "153",
    pages = "1",
    year = "1968"
}

@article{PhysRevA.30.1175,
  title = {Two-photon decay of hydrogenic atoms},
  author = {Tung, J. H. and Salamo, X. M. and Chan, F. T.},
  journal = {Phys. Rev. A},
  volume = {30},
  issue = {3},
  pages = {1175--1184},
  numpages = {0},
  year = {1984},
  month = {Sep},
  publisher = {American Physical Society},
  doi = {10.1103/PhysRevA.30.1175},
  url = {https://link.aps.org/doi/10.1103/PhysRevA.30.1175}
}

@article{PhysRevD.110.123506,
  title = {Primordial magnetic fields in light of dark ages global 21-cm signal},
  author = {Mohapatra, Vivekanand and Nayak, Alekha C. and Natwariya, Pravin Kumar},
  journal = {Phys. Rev. D},
  volume = {110},
  issue = {12},
  pages = {123506},
  numpages = {13},
  year = {2024},
  month = {Dec},
  publisher = {American Physical Society},
  doi = {10.1103/PhysRevD.110.123506},
  url = {https://link.aps.org/doi/10.1103/PhysRevD.110.123506}
}

@article{Short:2019twc,
    author = "Short, Katie and Bernal, Jos\'e Luis and Raccanelli, Alvise and Verde, Licia and Chluba, Jens",
    title = "{Enlightening the dark ages with dark matter}",
    eprint = "1912.07409",
    archivePrefix = "arXiv",
    primaryClass = "astro-ph.CO",
    doi = "10.1088/1475-7516/2020/07/020",
    journal = "JCAP",
    volume = "07",
    pages = "020",
    year = "2020"
}

@article{Mittal:2020kjs,
    author = "Mittal, Shikhar and Kulkarni, Girish",
    title = "{Ly \ensuremath{\alpha} coupling and heating at cosmic dawn}",
    eprint = "2009.10746",
    archivePrefix = "arXiv",
    primaryClass = "astro-ph.CO",
    doi = "10.1093/mnras/staa3811",
    journal = "Mon. Not. Roy. Astron. Soc.",
    volume = "503",
    number = "3",
    pages = "4264--4275",
    year = "2021"
}

@article{Hirata:2005mz,
    author = "Hirata, Christopher M.",
    title = "{Wouthuysen-Field coupling strength and application to high-redshift 21 cm radiation}",
    eprint = "astro-ph/0507102",
    archivePrefix = "arXiv",
    doi = "10.1111/j.1365-2966.2005.09949.x",
    journal = "Mon. Not. Roy. Astron. Soc.",
    volume = "367",
    pages = "259--274",
    year = "2006"
}

@ARTICLE{Pritchard:2011xb,
    author = "Pritchard, Jonathan R. and Loeb, Abraham",
    title = "{21-cm cosmology}",
    eprint = "1109.6012",
    archivePrefix = "arXiv",
    primaryClass = "astro-ph.CO",
    doi = "10.1088/0034-4885/75/8/086901",
    journal = "Rept. Prog. Phys.",
    volume = "75",
    pages = "086901",
    year = "2012"
}

@article{DAYAL20181,
title = {Early galaxy formation and its large-scale effects},
journal = {Physics Reports},
volume = {780-782},
pages = {1-64},
year = {2018},
note = {Early galaxy formation and its large-scale effects},
issn = {0370-1573},
doi = {https://doi.org/10.1016/j.physrep.2018.10.002},
url = {https://www.sciencedirect.com/science/article/pii/S0370157318302266},
author = {Pratika Dayal and Andrea Ferrara}
}

@article{Diemer:2017bwl,
    author = "Diemer, Benedikt",
    title = "{COLOSSUS: A python toolkit for cosmology, large-scale structure, and dark matter halos}",
    eprint = "1712.04512",
    archivePrefix = "arXiv",
    primaryClass = "astro-ph.CO",
    doi = "10.3847/1538-4365/aaee8c",
    journal = "Astrophys. J. Suppl.",
    volume = "239",
    number = "2",
    pages = "35",
    year = "2018"
}

@article{Barkana:2000fd,
    author = "Barkana, Rennan and Loeb, Abraham",
    title = "{In the beginning: The First sources of light and the reionization of the Universe}",
    eprint = "astro-ph/0010468",
    archivePrefix = "arXiv",
    doi = "10.1016/S0370-1573(01)00019-9",
    journal = "Phys. Rept.",
    volume = "349",
    pages = "125--238",
    year = "2001"
}

@article{Press:1973iz,
    author = "Press, William H. and Schechter, Paul",
    title = "{Formation of galaxies and clusters of galaxies by selfsimilar gravitational condensation}",
    doi = "10.1086/152650",
    journal = "Astrophys. J.",
    volume = "187",
    pages = "425--438",
    year = "1974"
}

@article{Barkana:2004vb,
    author = "Barkana, Rennan and Loeb, Abraham",
    title = "{Detecting the earliest galaxies through two new sources of 21cm fluctuations}",
    eprint = "astro-ph/0410129",
    archivePrefix = "arXiv",
    doi = "10.1086/429954",
    journal = "Astrophys. J.",
    volume = "626",
    pages = "1--11",
    year = "2005"
}

@ARTICLE{1952AJ.....57R..31W,
       author = {{Wouthuysen}, S.~A.},
        title = "{On the excitation mechanism of the 21-cm (radio-frequency) interstellar hydrogen emission line.}",
      journal = "Astrophys. J.",
         year = 1952,
        month = jan,
       volume = {57},
        pages = {31-32},
          doi = {10.1086/106661}
}

@ARTICLE{1959ApJ...129..536F,
       author = {{Field}, George B.},
        title = "{The Spin Temperature of Intergalactic Neutral Hydrogen.}",
      journal = "Astrophys. J.",
         year = 1959,
        month = may,
       volume = {129},
        pages = {536},
          doi = {10.1086/146653}
}

@ARTICLE{1958PIRE...46..240F,
       author = {{Field}, George B.},
        title = "{Excitation of the Hydrogen 21-CM Line}",
      journal = {Proceedings of the IRE},
         year = 1958,
        month = jan,
       volume = {46},
        pages = {240-250},
          doi = {10.1109/JRPROC.1958.286741}
      
}

@ARTICLE{Venumadhav:2018uwn,
    author = "Venumadhav, Tejaswi and Dai, Liang and Kaurov, Alexander and Zaldarriaga, Matias",
    title = "{Heating of the intergalactic medium by the cosmic microwave background during cosmic dawn}",
    eprint = "1804.02406",
    archivePrefix = "arXiv",
    primaryClass = "astro-ph.CO",
    doi = "10.1103/PhysRevD.98.103513",
    journal = "Phys. Rev. D",
    volume = "98",
    number = "10",
    pages = "103513",
    year = "2018"
}

@article{Acharya:2022txp,
    author = "Acharya, Sandeep Kumar and Dhandha, Jiten and Chluba, Jens",
    title = "{Can accreting primordial black holes explain the excess radio background?}",
    eprint = "2208.03816",
    archivePrefix = "arXiv",
    primaryClass = "astro-ph.CO",
    doi = "10.1093/mnras/stac2739",
    journal = "Mon. Not. Roy. Astron. Soc.",
    volume = "517",
    number = "2",
    pages = "2454--2461",
    year = "2022"
}

@article{Cyr:2023iwu,
    author = "Cyr, Bryce and Chluba, Jens and Acharya, Sandeep Kumar",
    title = "{Constraints on the spectral signatures of superconducting cosmic strings}",
    eprint = "2305.09816",
    archivePrefix = "arXiv",
    primaryClass = "astro-ph.CO",
    doi = "10.1093/mnras/stad2457",
    journal = "Mon. Not. Roy. Astron. Soc.",
    volume = "525",
    number = "2",
    pages = "2632--2653",
    year = "2023"
}

@article{Furlanetto:2006tf,
    author = "Furlanetto, Steven",
    title = "{The Global 21 Centimeter Background from High Redshifts}",
    eprint = "astro-ph/0604040",
    archivePrefix = "arXiv",
    doi = "10.1111/j.1365-2966.2006.10725.x",
    journal = "Mon. Not. Roy. Astron. Soc.",
    volume = "371",
    pages = "867--878",
    year = "2006"
}

@article{Mesinger:2010ne,
    author = "Mesinger, Andrei and Furlanetto, Steven and Cen, Renyue",
    title = "{21cmFAST: A Fast, Semi-Numerical Simulation of the High-Redshift 21-cm Signal}",
    eprint = "1003.3878",
    archivePrefix = "arXiv",
    primaryClass = "astro-ph.CO",
    doi = "10.1111/j.1365-2966.2010.17731.x",
    journal = "Mon. Not. Roy. Astron. Soc.",
    volume = "411",
    pages = "955",
    year = "2011"
}

@article{Furlanetto:2006jb,
    author = "Furlanetto, Steven and Oh, S. Peng and Briggs, Frank",
    title = "{Cosmology at Low Frequencies: The 21 cm Transition and the High-Redshift Universe}",
    eprint = "astro-ph/0608032",
    archivePrefix = "arXiv",
    doi = "10.1016/j.physrep.2006.08.002",
    journal = "Phys. Rept.",
    volume = "433",
    pages = "181--301",
    year = "2006"
}

@article{10.1111/j.1365-2966.2010.17940.x,
    author = {Chluba, J. and Thomas, R. M.},
    title = "{Towards a complete treatment of the cosmological recombination problem}",
    journal = {Monthly Notices of the Royal Astronomical Society},
    volume = {412},
    number = {2},
    pages = {748-764},
    year = {2011},
    month = {03},
    issn = {0035-8711},
    doi = {10.1111/j.1365-2966.2010.17940.x},
    url = {https://doi.org/10.1111/j.1365-2966.2010.17940.x}
}

@article{10.1111/j.1365-2966.2010.16940.x,
    author = {Chluba, J. and Vasil, G. M. and Dursi, L. J.},
    title = "{Recombinations to the Rydberg states of hydrogen and their effect during the cosmological recombination epoch}",
    journal = {Monthly Notices of the Royal Astronomical Society},
    volume = {407},
    number = {1},
    pages = {599-612},
    year = {2010},
    month = {08},
    issn = {0035-8711},
    doi = {10.1111/j.1365-2966.2010.16940.x},
    url = {https://doi.org/10.1111/j.1365-2966.2010.16940.x}
}

@article{Tseliakhovich:2010bj,
    author = "Tseliakhovich, Dmitriy and Hirata, Christopher",
    title = "{Relative velocity of dark matter and baryonic fluids and the formation of the first structures}",
    eprint = "1005.2416",
    archivePrefix = "arXiv",
    primaryClass = "astro-ph.CO",
    doi = "10.1103/PhysRevD.82.083520",
    journal = "Phys. Rev. D",
    volume = "82",
    pages = "083520",
    year = "2010"
}

@article{Scheck:2014cba,
    author = "Scheck, M. and others",
    editor = "Stoyanov, C. and Dimitrova, S.",
    title = "{Determination of the B(E3, 0(+) -{\ensuremath{>}} 3(-))-excitation strength in octupole-correlated nuclei near A approx 224 by the means of Coulomb excitation at REX-ISOLDE}",
    doi = "10.1088/1742-6596/533/1/012007",
    journal = "J. Phys. Conf. Ser.",
    volume = "533",
    pages = "012007",
    year = "2014"
}

@article{Hu:1994uz,
    author = "Hu, Wayne and Sugiyama, Naoshi",
    title = "{Anisotropies in the cosmic microwave background: An Analytic approach}",
    eprint = "astro-ph/9407093",
    archivePrefix = "arXiv",
    doi = "10.1086/175624",
    journal = "Astrophys. J.",
    volume = "444",
    pages = "489--506",
    year = "1995"
}

@article{Ma:1995ey,
    author = "Ma, Chung-Pei and Bertschinger, Edmund",
    title = "{Cosmological perturbation theory in the synchronous and conformal Newtonian gauges}",
    eprint = "astro-ph/9506072",
    archivePrefix = "arXiv",
    doi = "10.1086/176550",
    journal = "Astrophys. J.",
    volume = "455",
    pages = "7--25",
    year = "1995"
}

@article{Silk:1967kq,
    author = "Silk, Joseph",
    title = "{Cosmic black body radiation and galaxy formation}",
    doi = "10.1086/149449",
    journal = "Astrophys. J.",
    volume = "151",
    pages = "459--471",
    year = "1968"
}

@article{Barkana:2010zq,
    author = "Barkana, Rennan and Loeb, Abraham",
    title = "{Scale-Dependent Bias of Galaxies from Baryonic Acoustic Oscillations}",
    eprint = "1009.1393",
    archivePrefix = "arXiv",
    primaryClass = "astro-ph.CO",
    doi = "10.1111/j.1365-2966.2011.18922.x",
    journal = "Mon. Not. Roy. Astron. Soc.",
    volume = "415",
    pages = "3113",
    year = "2011"
}

@article{Ostriker:1998fa,
    author = "Ostriker, Eve C.",
    title = "{Dynamical friction in a gaseous medium}",
    eprint = "astro-ph/9810324",
    archivePrefix = "arXiv",
    doi = "10.1086/306858",
    journal = "Astrophys. J.",
    volume = "513",
    pages = "252",
    year = "1999"
}

@ARTICLE{1986A&A...164..337J,
       author = {{Just}, A. and {Kegel}, W.~H. and {Deiss}, B.~M.},
        title = "{Dynamical friction between the ISM and the system of stars}",
      journal = {Astron. Astrophys.},
         year = 1986,
        month = aug,
       volume = {164},
       number = {2},
        pages = {337-341},
       adsurl = {https://ui.adsabs.harvard.edu/abs/1986A&A...164..337J}
}

@ARTICLE{1990A&A...232..447J,
       author = {{Just}, A. and {Kegel}, W.~H.},
        title = "{Spatial structure and excitation timescales of fluctuations in the interstellar medium induced by the system of stars.}",
      journal = {Astron. Astrophys.},
         year = 1990,
        month = jun,
       volume = {232},
        pages = {447},
       adsurl = {https://ui.adsabs.harvard.edu/abs/1990A&A...232..447J}
}

@article{Kim:2025gck,
    author = "Kim, TaeHun and Lu, Philip and Takhistov, Volodymyr",
    title = "{Unified gas heating constraints on extended dark matter compact objects}",
    eprint = "2508.18344",
    archivePrefix = "arXiv",
    primaryClass = "hep-ph",
    reportNumber = "KEK-QUP-2025-0018, KEK-TH-2748",
    doi = "10.1088/1475-7516/2026/01/028",
    journal = "JCAP",
    volume = "01",
    pages = "028",
    year = "2026"
}

@article{Suzuguchi:2024btk,
    author = "Suzuguchi, Tomoya and Sugimura, Kazuyuki and Hosokawa, Takashi and Matsumoto, Tomoaki",
    title = "{Gas Dynamical Friction on Accreting Objects}",
    eprint = "2401.13032",
    archivePrefix = "arXiv",
    primaryClass = "astro-ph.GA",
    doi = "10.3847/1538-4357/ad34af",
    journal = "Astrophys. J.",
    volume = "966",
    number = "1",
    pages = "7",
    year = "2024"
}

@article{Bondi:1944rnk,
    author = "Bondi, H. and Hoyle, F.",
    title = "{On the Mechanism of Accretion by Stars}",
    doi = "10.1093/mnras/104.5.273",
    journal = "Mon. Not. Roy. Astron. Soc.",
    volume = "104",
    number = "5",
    pages = "273--282",
    year = "1944"
}

@article{Hoyle_Lyttleton_1939, 
title={The effect of interstellar matter on climatic variation}, volume={35}, 
DOI={10.1017/S0305004100021150}, 
number={3}, 
journal={Mathematical Proceedings of the Cambridge Philosophical Society}, 
author={Hoyle, F. and Lyttleton, R. A.}, 
year={1939},
pages={405–415}}

@article{Conroy:2007ps,
    author = "Conroy, Charlie and Ostriker, Jeremiah P.",
    title = "{Thermal Balance in the Intracluster Medium: Is AGN Feedback Necessary?}",
    eprint = "0712.0824",
    archivePrefix = "arXiv",
    primaryClass = "astro-ph",
    doi = "10.1086/587861",
    journal = "Astrophys. J.",
    volume = "681",
    pages = "151",
    year = "2008"
}

@article{Kim:2005ma,
    author = "Kim, Woong-Tae and El-Zant, Amr A. and Kamionkowski, Marc",
    title = "{Dynamical friction and cooling flows in galaxy clusters}",
    eprint = "astro-ph/0506579",
    archivePrefix = "arXiv",
    doi = "10.1086/432976",
    journal = "Astrophys. J.",
    volume = "632",
    pages = "157--168",
    year = "2005"
}

@article{Carr:2017jsz,
    author = {Carr, Bernard and Raidal, Martti and Tenkanen, Tommi and Vaskonen, Ville and Veerm{\"a}e, Hardi},
    title = "{Primordial black hole constraints for extended mass functions}",
    eprint = "1705.05567",
    archivePrefix = "arXiv",
    primaryClass = "astro-ph.CO",
    doi = "10.1103/PhysRevD.96.023514",
    journal = "Phys. Rev. D",
    volume = "96",
    number = "2",
    pages = "023514",
    year = "2017"
}

@article{Kannike:2017bxn,
    author = {Kannike, Kristjan and Marzola, Luca and Raidal, Martti and Veerm{\"a}e, Hardi},
    title = "{Single Field Double Inflation and Primordial Black Holes}",
    eprint = "1705.06225",
    archivePrefix = "arXiv",
    primaryClass = "astro-ph.CO",
    doi = "10.1088/1475-7516/2017/09/020",
    journal = "JCAP",
    volume = "09",
    pages = "020",
    year = "2017"
}

@article{Green:2016xgy,
    author = "Green, Anne M.",
    title = "{Microlensing and dynamical constraints on primordial black hole dark matter with an extended mass function}",
    eprint = "1609.01143",
    archivePrefix = "arXiv",
    primaryClass = "astro-ph.CO",
    doi = "10.1103/PhysRevD.94.063530",
    journal = "Phys. Rev. D",
    volume = "94",
    number = "6",
    pages = "063530",
    year = "2016"
}

@article{Kuhnel:2017pwq,
    author = {K{\"u}hnel, Florian and Freese, Katherine},
    title = "{Constraints on Primordial Black Holes with Extended Mass Functions}",
    eprint = "1701.07223",
    archivePrefix = "arXiv",
    primaryClass = "astro-ph.CO",
    doi = "10.1103/PhysRevD.95.083508",
    journal = "Phys. Rev. D",
    volume = "95",
    number = "8",
    pages = "083508",
    year = "2017"
}

@article{Dolgov:1992pu,
    author = "Dolgov, Alexandre and Silk, Joseph",
    title = "{Baryon isocurvature fluctuations at small scales and baryonic dark matter}",
    reportNumber = "CFPA-TH-92-04",
    doi = "10.1103/PhysRevD.47.4244",
    journal = "Phys. Rev. D",
    volume = "47",
    pages = "4244--4255",
    year = "1993"
}

@article{Yokoyama:1998xd,
    author = "Yokoyama, Jun'ichi",
    title = "{Cosmological constraints on primordial black holes produced in the near critical gravitational collapse}",
    eprint = "gr-qc/9804041",
    archivePrefix = "arXiv",
    reportNumber = "YITP-98-23, SU-ITP-98-18",
    doi = "10.1103/PhysRevD.58.107502",
    journal = "Phys. Rev. D",
    volume = "58",
    pages = "107502",
    year = "1998"
}

@article{Niemeyer:1999ak,
    author = "Niemeyer, Jens C. and Jedamzik, K.",
    title = "{Dynamics of primordial black hole formation}",
    eprint = "astro-ph/9901292",
    archivePrefix = "arXiv",
    doi = "10.1103/PhysRevD.59.124013",
    journal = "Phys. Rev. D",
    volume = "59",
    pages = "124013",
    year = "1999"
}

@article{Musco:2012au,
    author = "Musco, Ilia and Miller, John C.",
    title = "{Primordial black hole formation in the early universe: critical behaviour and self-similarity}",
    eprint = "1201.2379",
    archivePrefix = "arXiv",
    primaryClass = "gr-qc",
    doi = "10.1088/0264-9381/30/14/145009",
    journal = "Class. Quant. Grav.",
    volume = "30",
    pages = "145009",
    year = "2013"
}

@article{Carr:2016hva,
    author = "Carr, B. J. and Kohri, Kazunori and Sendouda, Yuuiti and Yokoyama, Jun'ichi",
    title = "{Constraints on primordial black holes from the Galactic gamma-ray background}",
    eprint = "1604.05349",
    archivePrefix = "arXiv",
    primaryClass = "astro-ph.CO",
    reportNumber = "RESCEU-16-16, KEK-TH-1895, KEK-COSMO-193",
    doi = "10.1103/PhysRevD.94.044029",
    journal = "Phys. Rev. D",
    volume = "94",
    number = "4",
    pages = "044029",
    year = "2016"
}

@article{Mohapatra:2025qpz,
    author = "Mohapatra, Vivekanand",
    title = "{Cosmological bounds on dark matter annihilation using dark ages 21-cm signal}",
    eprint = "2506.20648",
    archivePrefix = "arXiv",
    primaryClass = "astro-ph.CO",
    doi = "10.1016/j.dark.2025.102145",
    journal = "Phys. Dark Univ.",
    volume = "50",
    pages = "102145",
    year = "2025"
}

@article{Geller:1989da,
    author = "Geller, Margaret J. and Huchra, John P.",
    title = "{Mapping the universe}",
    doi = "10.1126/science.246.4932.897",
    journal = "Science",
    volume = "246",
    pages = "897--903",
    year = "1989"
}

@article{2DFGRS:2001zay,
    author = "Colless, Matthew and others",
    collaboration = "2DFGRS",
    title = "{The 2dF Galaxy Redshift Survey: Spectra and redshifts}",
    eprint = "astro-ph/0106498",
    archivePrefix = "arXiv",
    doi = "10.1046/j.1365-8711.2001.04902.x",
    journal = "Mon. Not. Roy. Astron. Soc.",
    volume = "328",
    pages = "1039",
    year = "2001"
}

@article{SDSS:2000hjo,
    author = "York, Donald G. and others",
    collaboration = "SDSS",
    title = "{The Sloan Digital Sky Survey: Technical Summary}",
    eprint = "astro-ph/0006396",
    archivePrefix = "arXiv",
    reportNumber = "FERMILAB-PUB-01-319-A",
    doi = "10.1086/301513",
    journal = "Astron. J.",
    volume = "120",
    pages = "1579--1587",
    year = "2000"
}

@article{Springel:2005nw,
    author = "Springel, Volker and others",
    title = "{Simulating the joint evolution of quasars, galaxies and their large-scale distribution}",
    eprint = "astro-ph/0504097",
    archivePrefix = "arXiv",
    doi = "10.1038/nature03597",
    journal = "Nature",
    volume = "435",
    pages = "629--636",
    year = "2005"
}

@article{Chatterjee:2019jts,
    author = "Chatterjee, Atrideb and Dayal, Pratika and Choudhury, Tirthankar Roy and Hutter, Anne",
    title = "{Ruling out 3 keV warm dark matter using 21 cm EDGES data}",
    eprint = "1902.09562",
    archivePrefix = "arXiv",
    primaryClass = "astro-ph.CO",
    doi = "10.1093/mnras/stz1444",
    journal = "Mon. Not. Roy. Astron. Soc.",
    volume = "487",
    number = "3",
    pages = "3560--3567",
    year = "2019"
}

@article{Matos:1998vk,
    author = "Matos, Tonatiuh and Guzman, Francisco Siddhartha",
    title = "{Scalar fields as dark matter in spiral galaxies}",
    eprint = "gr-qc/9810028",
    archivePrefix = "arXiv",
    reportNumber = "CINVESTAV-98-22",
    doi = "10.1088/0264-9381/17/1/102",
    journal = "Class. Quant. Grav.",
    volume = "17",
    pages = "L9--L16",
    year = "2000"
}

@article{Hu:2000ke,
    author = "Hu, Wayne and Barkana, Rennan and Gruzinov, Andrei",
    title = "{Cold and fuzzy dark matter}",
    eprint = "astro-ph/0003365",
    archivePrefix = "arXiv",
    doi = "10.1103/PhysRevLett.85.1158",
    journal = "Phys. Rev. Lett.",
    volume = "85",
    pages = "1158--1161",
    year = "2000"
}

@article{Arbey:2001qi,
    author = "Arbey, Alexandre and Lesgourgues, Julien and Salati, Pierre",
    title = "{Quintessential haloes around galaxies}",
    eprint = "astro-ph/0105564",
    archivePrefix = "arXiv",
    reportNumber = "LAPTH-850-01",
    doi = "10.1103/PhysRevD.64.123528",
    journal = "Phys. Rev. D",
    volume = "64",
    pages = "123528",
    year = "2001"
}

@article{Hui:2016ltb,
    author = "Hui, Lam and Ostriker, Jeremiah P. and Tremaine, Scott and Witten, Edward",
    title = "{Ultralight scalars as cosmological dark matter}",
    eprint = "1610.08297",
    archivePrefix = "arXiv",
    primaryClass = "astro-ph.CO",
    doi = "10.1103/PhysRevD.95.043541",
    journal = "Phys. Rev. D",
    volume = "95",
    number = "4",
    pages = "043541",
    year = "2017"
}

@article{Boyle:2001du,
    author = "Boyle, Latham A. and Caldwell, Robert R. and Kamionkowski, Marc",
    title = "{Spintessence! New models for dark matter and dark energy}",
    eprint = "astro-ph/0105318",
    archivePrefix = "arXiv",
    doi = "10.1016/S0370-2693(02)02590-X",
    journal = "Phys. Lett. B",
    volume = "545",
    pages = "17--22",
    year = "2002"
}

@article{Carlson:1992fn,
    author = "Carlson, Eric D. and Machacek, Marie E. and Hall, Lawrence J.",
    title = "{Self-interacting dark matter}",
    reportNumber = "HUTP-91-A066, LBL-32016, UCB-92-06, NUB-3042-92-TH",
    doi = "10.1086/171833",
    journal = "Astrophys. J.",
    volume = "398",
    pages = "43--52",
    year = "1992"
}

@article{Spergel:1999mh,
    author = "Spergel, David N. and Steinhardt, Paul J.",
    title = "{Observational evidence for selfinteracting cold dark matter}",
    eprint = "astro-ph/9909386",
    archivePrefix = "arXiv",
    doi = "10.1103/PhysRevLett.84.3760",
    journal = "Phys. Rev. Lett.",
    volume = "84",
    pages = "3760--3763",
    year = "2000"
}

@article{Bhatt:2019qbq,
    author = "Bhatt, Jitesh R. and Mishra, Arvind Kumar and Nayak, Alekha C.",
    title = "{Viscous dark matter and 21 cm cosmology}",
    eprint = "1901.08451",
    archivePrefix = "arXiv",
    primaryClass = "astro-ph.CO",
    doi = "10.1103/PhysRevD.100.063539",
    journal = "Phys. Rev. D",
    volume = "100",
    number = "6",
    pages = "063539",
    year = "2019"
}

@article{Green:2020jor,
    author = "Green, Anne M. and Kavanagh, Bradley J.",
    title = "{Primordial Black Holes as a dark matter candidate}",
    eprint = "2007.10722",
    archivePrefix = "arXiv",
    primaryClass = "astro-ph.CO",
    doi = "10.1088/1361-6471/abc534",
    journal = "J. Phys. G",
    volume = "48",
    number = "4",
    pages = "043001",
    year = "2021"
}

@article{Witten:1984rs,
    author = "Witten, Edward",
    title = "{Cosmic Separation of Phases}",
    reportNumber = "PRINT-84-0400 (IAS,PRINCETON)",
    doi = "10.1103/PhysRevD.30.272",
    journal = "Phys. Rev. D",
    volume = "30",
    pages = "272--285",
    year = "1984"
}

@article{Lee:1991ax,
    author = "Lee, T. D. and Pang, Y.",
    editor = "Ren, Hai-Cang and Pang, Yang",
    title = "{Nontopological solitons}",
    reportNumber = "CU-TP-506",
    doi = "10.1016/0370-1573(92)90064-7",
    journal = "Phys. Rept.",
    volume = "221",
    pages = "251--350",
    year = "1992"
}

@article{Kolb:1993zz,
    author = "Kolb, Edward W. and Tkachev, Igor I.",
    title = "{Axion miniclusters and Bose stars}",
    eprint = "hep-ph/9303313",
    archivePrefix = "arXiv",
    reportNumber = "FERMILAB-PUB-93-066-A",
    doi = "10.1103/PhysRevLett.71.3051",
    journal = "Phys. Rev. Lett.",
    volume = "71",
    pages = "3051--3054",
    year = "1993"
}

@article{Coleman:1985ki,
    author = "Coleman, Sidney R.",
    title = "{Q-balls}",
    reportNumber = "HUTP-85/A050",
    doi = "10.1016/0550-3213(86)90520-1",
    journal = "Nucl. Phys. B",
    volume = "262",
    number = "2",
    pages = "263",
    year = "1985",
    note = "[Addendum: Nucl.Phys.B 269, 744 (1986)]"
}

@article{Grabowska:2018lnd,
    author = "Grabowska, Dorota M. and Melia, Tom and Rajendran, Surjeet",
    title = "{Detecting Dark Blobs}",
    eprint = "1807.03788",
    archivePrefix = "arXiv",
    primaryClass = "hep-ph",
    doi = "10.1103/PhysRevD.98.115020",
    journal = "Phys. Rev. D",
    volume = "98",
    number = "11",
    pages = "115020",
    year = "2018"
}

@article{Carr:2020xqk,
    author = "Carr, Bernard and Kuhnel, Florian",
    title = "{Primordial Black Holes as Dark Matter: Recent Developments}",
    eprint = "2006.02838",
    archivePrefix = "arXiv",
    primaryClass = "astro-ph.CO",
    doi = "10.1146/annurev-nucl-050520-125911",
    journal = "Ann. Rev. Nucl. Part. Sci.",
    volume = "70",
    pages = "355--394",
    year = "2020"
}

@article{Ricotti:2007au,
    author = "Ricotti, Massimo and Ostriker, Jeremiah P. and Mack, Katherine J.",
    title = "{Effect of Primordial Black Holes on the Cosmic Microwave Background and Cosmological Parameter Estimates}",
    eprint = "0709.0524",
    archivePrefix = "arXiv",
    primaryClass = "astro-ph",
    doi = "10.1086/587831",
    journal = "Astrophys. J.",
    volume = "680",
    pages = "829",
    year = "2008"
}

@article{Ali-Haimoud:2016mbv,
    author = {Ali-Ha{\"\i}moud, Yacine and Kamionkowski, Marc},
    title = "{Cosmic microwave background limits on accreting primordial black holes}",
    eprint = "1612.05644",
    archivePrefix = "arXiv",
    primaryClass = "astro-ph.CO",
    doi = "10.1103/PhysRevD.95.043534",
    journal = "Phys. Rev. D",
    volume = "95",
    number = "4",
    pages = "043534",
    year = "2017"
}

@article{Ruffini:1969qy,
    author = "Ruffini, Remo and Bonazzola, Silvano",
    title = "{Systems of selfgravitating particles in general relativity and the concept of an equation of state}",
    doi = "10.1103/PhysRev.187.1767",
    journal = "Phys. Rev.",
    volume = "187",
    pages = "1767--1783",
    year = "1969"
}

@article{Liebling:2012fv,
    author = "Liebling, Steven L. and Palenzuela, Carlos",
    title = "{Dynamical boson stars}",
    eprint = "1202.5809",
    archivePrefix = "arXiv",
    primaryClass = "gr-qc",
    doi = "10.1007/s41114-023-00043-4",
    journal = "Living Rev. Rel.",
    volume = "26",
    number = "1",
    pages = "1",
    year = "2023"
}

@article{Chang:2024fol,
    author = "Chang, Jae Hyeok and Fox, Patrick J. and Xiao, Huangyu",
    title = "{Axion stars: mass functions and constraints}",
    eprint = "2406.09499",
    archivePrefix = "arXiv",
    primaryClass = "hep-ph",
    reportNumber = "FERMILAB-PUB-24-0295-T",
    doi = "10.1088/1475-7516/2024/08/023",
    journal = "JCAP",
    volume = "08",
    pages = "023",
    year = "2024"
}

@article{Macho:2000nvd,
    author = "Allsman, R. A. and others",
    collaboration = "Macho",
    title = "{MACHO project limits on black hole dark matter in the 1-30 solar mass range}",
    eprint = "astro-ph/0011506",
    archivePrefix = "arXiv",
    doi = "10.1086/319636",
    journal = "Astrophys. J. Lett.",
    volume = "550",
    pages = "L169",
    year = "2001"
}

@article{Niikura:2019kqi,
    author = "Niikura, Hiroko and Takada, Masahiro and Yokoyama, Shuichiro and Sumi, Takahiro and Masaki, Shogo",
    title = "{Constraints on Earth-mass primordial black holes from OGLE 5-year microlensing events}",
    eprint = "1901.07120",
    archivePrefix = "arXiv",
    primaryClass = "astro-ph.CO",
    doi = "10.1103/PhysRevD.99.083503",
    journal = "Phys. Rev. D",
    volume = "99",
    number = "8",
    pages = "083503",
    year = "2019"
}

@article{EROS-2:2006ryy,
    author = "Tisserand, P. and others",
    collaboration = "EROS-2",
    title = "{Limits on the Macho Content of the Galactic Halo from the EROS-2 Survey of the Magellanic Clouds}",
    eprint = "astro-ph/0607207",
    archivePrefix = "arXiv",
    doi = "10.1051/0004-6361:20066017",
    journal = "Astron. Astrophys.",
    volume = "469",
    pages = "387--404",
    year = "2007"
}

@article{Oguri:2017ock,
    author = "Oguri, Masamune and Diego, Jose M. and Kaiser, Nick and Kelly, Patrick L. and Broadhurst, Tom",
    title = "{Understanding caustic crossings in giant arcs: characteristic scales, event rates, and constraints on compact dark matter}",
    eprint = "1710.00148",
    archivePrefix = "arXiv",
    primaryClass = "astro-ph.CO",
    doi = "10.1103/PhysRevD.97.023518",
    journal = "Phys. Rev. D",
    volume = "97",
    number = "2",
    pages = "023518",
    year = "2018"
}

@article{Wilkinson:2001vv,
    author = "Wilkinson, P. N. and Henstock, D. R. and Browne, I. W. A. and Polatidis, A. G. and Augusto, P. and Readhead, A. C. S. and Pearson, T. J. and Xu, W. and Taylor, G. B. and Vermeulen, R. C.",
    title = "{Limits on the cosmological abundance of supermassive compact objects from a search for multiple imaging in compact radio sources}",
    eprint = "astro-ph/0101328",
    archivePrefix = "arXiv",
    doi = "10.1103/PhysRevLett.86.584",
    journal = "Phys. Rev. Lett.",
    volume = "86",
    pages = "584--587",
    year = "2001"
}

@article{Yoo:2003fr,
    author = "Yoo, Jaiyul and Chaname, Julio and Gould, Andrew",
    title = "{The end of the MACHO era: limits on halo dark matter from stellar halo wide binaries}",
    eprint = "astro-ph/0307437",
    archivePrefix = "arXiv",
    doi = "10.1086/380562",
    journal = "Astrophys. J.",
    volume = "601",
    pages = "311--318",
    year = "2004"
}

@article{Carr:2020gox,
    author = "Carr, Bernard and Kohri, Kazunori and Sendouda, Yuuiti and Yokoyama, Jun'ichi",
    title = "{Constraints on primordial black holes}",
    eprint = "2002.12778",
    archivePrefix = "arXiv",
    primaryClass = "astro-ph.CO",
    reportNumber = "RESCEU-03/20; KEK-Cosmo-249; KEK-TH-2199; IPMU20-0024",
    doi = "10.1088/1361-6633/ac1e31",
    journal = "Rept. Prog. Phys.",
    volume = "84",
    number = "11",
    pages = "116902",
    year = "2021"
}

@article{Moore:1993sv,
    author = "Moore, Ben",
    title = "{An Upper limit to the mass of black holes in the halo of our galaxy}",
    eprint = "astro-ph/9306004",
    archivePrefix = "arXiv",
    reportNumber = "PRINT-93-0458 (UC,BERKELEY)",
    doi = "10.1086/186967",
    journal = "Astrophys. J. Lett.",
    volume = "413",
    pages = "L93",
    year = "1993"
}

@article{Udalski2015OGLEIVFP,
  title={OGLE-IV: Fourth Phase of the Optical Gravitational Lensing Experiment},
  author={Andrzej Udalski and Michal K. Szyma'nski and G. Szyma'nski},
  journal={arXiv: Solar and Stellar Astrophysics},
  year={2015},
  url={https://api.semanticscholar.org/CorpusID:119180082}
}

@article{Blaineau:2022nhy,
    author = "Blaineau, T. and others",
    title = "{New limits from microlensing on Galactic black holes in the mass range 10 M{\ensuremath{\odot}} {\ensuremath{<}} M {\ensuremath{<}} 1000 M{\ensuremath{\odot}}}",
    eprint = "2202.13819",
    archivePrefix = "arXiv",
    primaryClass = "astro-ph.GA",
    doi = "10.1051/0004-6361/202243430",
    journal = "Astron. Astrophys.",
    volume = "664",
    pages = "A106",
    year = "2022"
}

@article{Bai:2020jfm,
    author = "Bai, Yang and Long, Andrew J. and Lu, Sida",
    title = "{Tests of Dark MACHOs: Lensing, Accretion, and Glow}",
    eprint = "2003.13182",
    archivePrefix = "arXiv",
    primaryClass = "astro-ph.CO",
    doi = "10.1088/1475-7516/2020/09/044",
    journal = "JCAP",
    volume = "09",
    pages = "044",
    year = "2020"
}

@article{Lu:2020bmd,
    author = "Lu, Philip and Takhistov, Volodymyr and Gelmini, Graciela B. and Hayashi, Kohei and Inoue, Yoshiyuki and Kusenko, Alexander",
    title = "{Constraining Primordial Black Holes with Dwarf Galaxy Heating}",
    eprint = "2007.02213",
    archivePrefix = "arXiv",
    primaryClass = "astro-ph.CO",
    reportNumber = "IPMU20-0076, RIKEN-iTHEMS-Report-20",
    doi = "10.3847/2041-8213/abdcb6",
    journal = "Astrophys. J. Lett.",
    volume = "908",
    number = "2",
    pages = "L23",
    year = "2021"
}

@article{Takhistov:2021aqx,
    author = "Takhistov, Volodymyr and Lu, Philip and Gelmini, Graciela B. and Hayashi, Kohei and Inoue, Yoshiyuki and Kusenko, Alexander",
    title = "{Interstellar gas heating by primordial black holes}",
    eprint = "2105.06099",
    archivePrefix = "arXiv",
    primaryClass = "astro-ph.GA",
    reportNumber = "IPMU21-0029, RIKEN-iTHEMS-Report-21",
    doi = "10.1088/1475-7516/2022/03/017",
    journal = "JCAP",
    volume = "03",
    number = "03",
    pages = "017",
    year = "2022"
}

@article{Brandt:2016aco,
    author = "Brandt, Timothy D.",
    title = "{Constraints on MACHO Dark Matter from Compact Stellar Systems in Ultra-Faint Dwarf Galaxies}",
    eprint = "1605.03665",
    archivePrefix = "arXiv",
    primaryClass = "astro-ph.GA",
    doi = "10.3847/2041-8205/824/2/L31",
    journal = "Astrophys. J. Lett.",
    volume = "824",
    number = "2",
    pages = "L31",
    year = "2016"
}

@article{Wadekar:2022ymq,
    author = "Wadekar, Digvijay and Wang, Zihui",
    title = "{Constraining axion and compact dark matter with interstellar medium heating}",
    eprint = "2211.07668",
    archivePrefix = "arXiv",
    primaryClass = "hep-ph",
    doi = "10.1103/PhysRevD.107.083011",
    journal = "Phys. Rev. D",
    volume = "107",
    number = "8",
    pages = "083011",
    year = "2023"
}

@article{Graham:2025opw,
    author = "Graham, Peter W. and Ramani, Harikrishnan and Ruhdorfer, Maximilian",
    title = "{Robust bounds on MACHOs from the faintest galaxies}",
    eprint = "2510.01310",
    archivePrefix = "arXiv",
    primaryClass = "hep-ph",
    doi = "10.1103/2dv9-xxtw",
    journal = "Phys. Rev. D",
    volume = "113",
    number = "2",
    pages = "023047",
    year = "2026"
}

@article{Singh:2021mxo,
    author = "Singh, Saurabh and Nambissan T., Jishnu and Subrahmanyan, Ravi and Udaya Shankar, N. and Girish, B. S. and Raghunathan, A. and Somashekar, R. and Srivani, K. S. and Sathyanarayana Rao, Mayuri",
    title = "{On the detection of a cosmic dawn signal in the radio background}",
    eprint = "2112.06778",
    archivePrefix = "arXiv",
    primaryClass = "astro-ph.CO",
    doi = "10.1038/s41550-022-01610-5",
    journal = "Nature Astron.",
    volume = "6",
    number = "5",
    pages = "607--617",
    year = "2022"
}

@INPROCEEDINGS{2019BAAS...51g.178B,
       author = {{Burns}, Jack and {Hallinan}, Gregg and {Lux}, Jim and {Romero-Wolf}, Andres and {Teitelbaum}, Lawrence and {Chang}, Tzu-Ching and {Kocz}, Jonathan and {Bowman}, Judd and {MacDowall}, Robert and {Kasper}, Justin and {Bradley}, Richard and {Anderson}, Marin and {Rapetti}, David},
        title = "{FARSIDE: A Low Radio Frequency Interferometric Array on the Lunar Farside}",
    booktitle = {Bulletin of the American Astronomical Society},
         year = {2019},
       volume = {51},
        month = {sep},
          eid = {178},
        pages = {178},
          doi = {10.48550/arXiv.1907.05407},
archivePrefix = {arXiv},
       eprint = {1907.05407},
 primaryClass = {astro-ph.IM},
       adsurl = {https://ui.adsabs.harvard.edu/abs/2019BAAS...51g.178B}
}

@article{10.1063/5.0043435,
    author = {Borade, R. and George, George N. and Gharpure, D. C.},
    title = {FPGA based data acquisition and processing system for space electric and magnetic sensors (SEAMS)},
    journal = {AIP Conference Proceedings},
    volume = {2335},
    number = {1},
    pages = {030005},
    year = {2021},
    month = {03},
    issn = {0094-243X},
    doi = {10.1063/5.0043435},
    url = {https://doi.org/10.1063/5.0043435},
}

@ARTICLE{2023arXiv230110345B,
       author = {{Bale}, Stuart D. and {Bassett}, Neil and {Burns}, Jack O. and {Dorigo Jones}, Johnny and {Goetz}, Keith and {Hellum-Bye}, Christian and {Hermann}, Sven and {Hibbard}, Joshua and {Maksimovic}, Milan and {McLean}, Ryan and {Monsalve}, Raul and {O'Connor}, Paul and {Parsons}, Aaron and {Pulupa}, Marc and {Pund}, Rugved and {Rapetti}, David and {Rotermund}, Kaja M. and {Saliwanchik}, Ben and {Slosar}, Anze and {Sundkvist}, David and {Suzuki}, Aritoki},
        title = "{LuSEE 'Night': The Lunar Surface Electromagnetics Experiment}",
      journal = {arXiv e-prints},
         year = 2023,
        month = jan,
          eid = {arXiv:2301.10345},
        pages = {arXiv:2301.10345},
          doi = {10.48550/arXiv.2301.10345},
archivePrefix = {arXiv},
       eprint = {2301.10345},
 primaryClass = {astro-ph.IM},
       adsurl = {https://ui.adsabs.harvard.edu/abs/2023arXiv230110345B}
}

@ARTICLE{2023ExA....56..741S,
       author = {{Sathyanarayana Rao}, Mayuri and {Singh}, Saurabh and {K.~S.}, Srivani and {B.~S.}, Girish and {Sathish}, Keerthipriya and {Somashekar}, R. and {Agaram}, Raghunathan and {Kavitha}, K. and {Vishwapriya}, Gautam and {Anand}, Ashish and {Udaya Shankar}, N. and {Seetha}, S.},
        title = "{PRATUSH experiment concept and design overview}",
      journal = {Experimental Astronomy},
         year = 2023,
        month = sep,
       volume = {56},
       number = {2-3},
        pages = {741-778},
          doi = {10.1007/s10686-023-09909-5},
archivePrefix = {arXiv},
       eprint = {2507.05654},
 primaryClass = {astro-ph.IM},
       adsurl = {https://ui.adsabs.harvard.edu/abs/2023ExA....56..741S}
}

@article{deLeraAcedo:2022kiu,
    author = "de Lera Acedo, E. and others",
    title = "{The REACH radiometer for detecting the 21-cm hydrogen signal from redshift z{\,}{\ensuremath{\approx}}{\,}7.5{\textendash}28}",
    eprint = "2210.07409",
    archivePrefix = "arXiv",
    primaryClass = "astro-ph.CO",
    doi = "10.1038/s41550-022-01817-6",
    journal = "Nature Astron.",
    volume = "6",
    number = "7",
    pages = "998",
    year = "2022"
}

@article{Burns:2020gfh,
    author = "Burns, Jack O.",
    title = "{Transformative Science from the Lunar Farside: Observations of the Dark Ages and Exoplanetary Systems at Low Radio Frequencies}",
    eprint = "2003.06881",
    archivePrefix = "arXiv",
    primaryClass = "astro-ph.IM",
    doi = "10.1098/rsta.2019.0564",
    journal = "Philosophical Transactions of the Royal Society A: Mathematical, Physical and Engineering Sciences",
    volume = "379",
    number = "2188",
    year = "2020"
}

@article{Rapetti:2019lmf,
    author = "Rapetti, David and Tauscher, Keith and Mirocha, Jordan and Burns, Jack O.",
    title = "{Global 21-cm Signal Extraction from Foreground and Instrumental Effects II: Efficient and Self-Consistent Technique for Constraining Nonlinear Signal Models}",
    eprint = "1912.02205",
    archivePrefix = "arXiv",
    primaryClass = "astro-ph.CO",
    doi = "10.3847/1538-4357/ab9b29",
    journal = "Astrophys. J.",
    volume = "897",
    number = "2",
    pages = "174",
    year = "2020"
}

@INPROCEEDINGS{10906958,
  author={Bale, Stuart D. and Bonnell, John W. and Burns, Jack O. and De Wit, Thierry Dudok and Fahs, Adam and Goetz, Keith and Hellum-Bye, Christian and Herrmann, Sven and Hibbard, Joshua and Li, Zack and Maksimovic, Milan and Malaspina, David and McLean, Ryan and Monsalve, Raul and O'Connor, Paul and Page, Brent and Parsons, Aaron and Pulupa, Marc and Pund, Rugved and Rapetti, David and Rotermund, Kaja M. and Saliwanchik, Ben and Sheppard, Dave and Slosar, Anže and Sundkvist, David and Suzuki, Aritoki and Yousuf, Fatima},
  booktitle={2025 United States National Committee of URSI National Radio Science Meeting (USNC-URSI NRSM)}, 
  title={Pioneering Ultra-Long-Wavelength Radio Science with LuSEE-Night}, 
  year={2025},
  volume={},
  number={},
  pages={45-45},
  doi={10.23919/USNC-URSINRSM66067.2025.10906958}}

@Article{Hunter:2007,
  Author    = {Hunter, J. D.},
  Title     = {Matplotlib: A 2D graphics environment},
  Journal   = {Computing in Science \& Engineering},
  Volume    = {9},
  Number    = {3},
  Pages     = {90--95},
  publisher = {IEEE COMPUTER SOC},
  doi       = {10.1109/MCSE.2007.55},
  year      = 2007
}

@ARTICLE{2020SciPy-NMeth,
  author  = {Virtanen, Pauli and Gommers, Ralf and Oliphant, Travis E. and
            Haberland, Matt and Reddy, Tyler and Cournapeau, David and
            Burovski, Evgeni and Peterson, Pearu and Weckesser, Warren and
            Bright, Jonathan and {van der Walt}, St{\'e}fan J. and
            Brett, Matthew and Wilson, Joshua and Millman, K. Jarrod and
            Mayorov, Nikolay and Nelson, Andrew R. J. and Jones, Eric and
            Kern, Robert and Larson, Eric and Carey, C J and
            Polat, {\.I}lhan and Feng, Yu and Moore, Eric W. and
            {VanderPlas}, Jake and Laxalde, Denis and Perktold, Josef and
            Cimrman, Robert and Henriksen, Ian and Quintero, E. A. and
            Harris, Charles R. and Archibald, Anne M. and
            Ribeiro, Ant{\^o}nio H. and Pedregosa, Fabian and
            {van Mulbregt}, Paul and {SciPy 1.0 Contributors}},
  title   = {{{SciPy} 1.0: Fundamental Algorithms for Scientific
            Computing in Python}},
  journal = {Nature Methods},
  year    = {2020},
  volume  = {17},
  pages   = {261--272},
  adsurl  = {https://rdcu.be/b08Wh},
  doi     = {10.1038/s41592-019-0686-2},
}

@Article{         harris2020array,
 title         = {Array programming with {NumPy}},
 author        = {Charles R. Harris and K. Jarrod Millman and St{\'{e}}fan J.
                 van der Walt and Ralf Gommers and Pauli Virtanen and David
                 Cournapeau and Eric Wieser and Julian Taylor and Sebastian
                 Berg and Nathaniel J. Smith and Robert Kern and Matti Picus
                 and Stephan Hoyer and Marten H. van Kerkwijk and Matthew
                 Brett and Allan Haldane and Jaime Fern{\'{a}}ndez del
                 R{\'{i}}o and Mark Wiebe and Pearu Peterson and Pierre
                 G{\'{e}}rard-Marchant and Kevin Sheppard and Tyler Reddy and
                 Warren Weckesser and Hameer Abbasi and Christoph Gohlke and
                 Travis E. Oliphant},
 year          = {2020},
 month         = sep,
 journal       = {Nature},
 volume        = {585},
 number        = {7825},
 pages         = {357--362},
 doi           = {10.1038/s41586-020-2649-2},
 publisher     = {Springer Science and Business Media {LLC}},
 url           = {https://doi.org/10.1038/s41586-020-2649-2}
}

@article{Leach:2001zf,
    author = "Leach, Samuel M and Sasaki, Misao and Wands, David and Liddle, Andrew R",
    title = "{Enhancement of superhorizon scale inflationary curvature perturbations}",
    eprint = "astro-ph/0101406",
    archivePrefix = "arXiv",
    doi = "10.1103/PhysRevD.64.023512",
    journal = "Phys. Rev. D",
    volume = "64",
    pages = "023512",
    year = "2001"
}

@article{Pajer:2013fsa,
    author = "Pajer, Enrico and Peloso, Marco",
    title = "{A review of Axion Inflation in the era of Planck}",
    eprint = "1305.3557",
    archivePrefix = "arXiv",
    primaryClass = "hep-th",
    doi = "10.1088/0264-9381/30/21/214002",
    journal = "Class. Quant. Grav.",
    volume = "30",
    pages = "214002",
    year = "2013"
}

@article{Hertzberg:2017dkh,
    author = "Hertzberg, Mark P. and Yamada, Masaki",
    title = "{Primordial Black Holes from Polynomial Potentials in Single Field Inflation}",
    eprint = "1712.09750",
    archivePrefix = "arXiv",
    primaryClass = "astro-ph.CO",
    doi = "10.1103/PhysRevD.97.083509",
    journal = "Phys. Rev. D",
    volume = "97",
    number = "8",
    pages = "083509",
    year = "2018"
}

@article{Mishra:2019pzq,
    author = "Mishra, Swagat S. and Sahni, Varun",
    title = "{Primordial Black Holes from a tiny bump/dip in the Inflaton potential}",
    eprint = "1911.00057",
    archivePrefix = "arXiv",
    primaryClass = "gr-qc",
    doi = "10.1088/1475-7516/2020/04/007",
    journal = "JCAP",
    volume = "04",
    pages = "007",
    year = "2020"
}

@article{Inomata:2021tpx,
    author = "Inomata, Keisuke and McDonough, Evan and Hu, Wayne",
    title = "{Amplification of primordial perturbations from the rise or fall of the inflaton}",
    eprint = "2110.14641",
    archivePrefix = "arXiv",
    primaryClass = "astro-ph.CO",
    doi = "10.1088/1475-7516/2022/02/031",
    journal = "JCAP",
    volume = "02",
    number = "02",
    pages = "031",
    year = "2022"
}

@article{Hooper:2023nnl,
    author = "Hooper, Dan and Ireland, Aurora and Krnjaic, Gordan and Stebbins, Albert",
    title = "{Supermassive primordial black holes from inflation}",
    eprint = "2308.00756",
    archivePrefix = "arXiv",
    primaryClass = "astro-ph.CO",
    reportNumber = "FERMILAB-PUB-23-390-T",
    doi = "10.1088/1475-7516/2024/04/021",
    journal = "JCAP",
    volume = "04",
    pages = "021",
    year = "2024"
}

@article{Erickcek:2011us,
    author = "Erickcek, Adrienne L. and Sigurdson, Kris",
    title = "{Reheating Effects in the Matter Power Spectrum and Implications for Substructure}",
    eprint = "1106.0536",
    archivePrefix = "arXiv",
    primaryClass = "astro-ph.CO",
    doi = "10.1103/PhysRevD.84.083503",
    journal = "Phys. Rev. D",
    volume = "84",
    pages = "083503",
    year = "2011"
}

@article{Fan:2014zua,
    author = {Fan, JiJi and {\"O}zsoy, Ogan and Watson, Scott},
    title = "{Nonthermal histories and implications for structure formation}",
    eprint = "1405.7373",
    archivePrefix = "arXiv",
    primaryClass = "hep-ph",
    doi = "10.1103/PhysRevD.90.043536",
    journal = "Phys. Rev. D",
    volume = "90",
    number = "4",
    pages = "043536",
    year = "2014"
}

@article{Erickcek:2015jza,
    author = "Erickcek, Adrienne L.",
    title = "{The Dark Matter Annihilation Boost from Low-Temperature Reheating}",
    eprint = "1504.03335",
    archivePrefix = "arXiv",
    primaryClass = "astro-ph.CO",
    doi = "10.1103/PhysRevD.92.103505",
    journal = "Phys. Rev. D",
    volume = "92",
    number = "10",
    pages = "103505",
    year = "2015"
}

@article{Dror:2017gjq,
    author = "Dror, Jeff A. and Kuflik, Eric and Melcher, Brandon and Watson, Scott",
    title = "{Concentrated dark matter: Enhanced small-scale structure from codecaying dark matter}",
    eprint = "1711.04773",
    archivePrefix = "arXiv",
    primaryClass = "hep-ph",
    doi = "10.1103/PhysRevD.97.063524",
    journal = "Phys. Rev. D",
    volume = "97",
    number = "6",
    pages = "063524",
    year = "2018"
}

@article{Kohri:2022wzp,
    author = "Kohri, Kazunori and Sekiguchi, Toyokazu and Wang, Sai",
    title = "{Cosmological 21-cm line observations to test scenarios of super-Eddington accretion on to black holes being seeds of high-redshifted supermassive black holes}",
    eprint = "2201.05300",
    archivePrefix = "arXiv",
    primaryClass = "astro-ph.CO",
    reportNumber = "KEK-Cosmo-0284, KEK-TH-2389",
    doi = "10.1103/PhysRevD.106.043539",
    journal = "Phys. Rev. D",
    volume = "106",
    number = "4",
    pages = "043539",
    year = "2022"
}

@article{Mroz:2024mse,
    author = "Mr{\'o}z, Przemek and others",
    title = "{No massive black holes in the Milky Way halo}",
    eprint = "2403.02386",
    archivePrefix = "arXiv",
    primaryClass = "astro-ph.GA",
    doi = "10.1038/s41586-024-07704-6",
    journal = "Nature",
    volume = "632",
    number = "8026",
    pages = "749--751",
    year = "2024"
}

@article{Mroz:2024wia,
    author = "Mr{\'o}z, Przemek and others",
    title = "{Limits on Planetary-mass Primordial Black Holes from the OGLE High-cadence Survey of the Magellanic Clouds}",
    eprint = "2410.06251",
    archivePrefix = "arXiv",
    primaryClass = "astro-ph.CO",
    doi = "10.3847/2041-8213/ad8e68",
    journal = "Astrophys. J. Lett.",
    volume = "976",
    number = "1",
    pages = "L19",
    year = "2024"
}

@article{Kim_2009,
doi = {10.1088/0004-637X/703/2/1278},
url = {https://doi.org/10.1088/0004-637X/703/2/1278},
year = {2009},
month = {sep},
publisher = {The American Astronomical Society},
volume = {703},
number = {2},
pages = {1278},
author = {Kim, Hyosun and Kim, Woong-Tae},
title = {NONLINEAR DYNAMICAL FRICTION IN A GASEOUS MEDIUM},
journal = {The Astrophysical Journal}
}
\end{document}